\documentclass[%
 reprint,
superscriptaddress,
 amsmath,amssymb,
 aps,
prb,
]{revtex4-2}

\usepackage{graphicx}% Include figure files
\usepackage{dcolumn}% Align table columns on decimal point
\usepackage{bm}% bold math
\usepackage{multirow}
\usepackage[abs]{overpic}
\usepackage{hyperref}
\hypersetup{hidelinks} 
\hypersetup{
 hypertex=true,
 colorlinks=true,
 urlcolor=blue,
 linkcolor=blue,
 anchorcolor=red,
 citecolor=blue
 }
\usepackage{threeparttable}
\usepackage{enumitem}
\usepackage{array}
\usepackage{booktabs}

\begin{document}

\preprint{APS/123-QED}

\title{
Mechanisms of Dual-Band Emission in Sb-Doped Rare-Earth Phosphates Revealed
}% Force line breaks with \\
\author{Ruijie Hao}
\affiliation{CAS Key Laboratory of Microscale Magnetic Resonance, and School of Physical Sciences, University of Science and Technology of China, Hefei 230026, China}
\affiliation{Anhui Province Key Laboratory of Scientific Instrument Development and Application, University of Science and Technology of China, Hefei 230026, China}
\affiliation{CAS Center for Excellence in Quantum Information and Quantum Physics, University of Science and Technology of China, Hefei 230026, China}
\author{Xin Zhao}
\affiliation{CAS Key Laboratory of Microscale Magnetic Resonance, and School of Physical Sciences, University of Science and Technology of China, Hefei 230026, China}
\affiliation{Anhui Province Key Laboratory of Scientific Instrument Development and Application, University of Science and Technology of China, Hefei 230026, China}
\affiliation{CAS Center for Excellence in Quantum Information and Quantum Physics, University of Science and Technology of China, Hefei 230026, China}
\author{Chang-Kui Duan}
\email{ckduan@ustc.edu.cn}
\affiliation{CAS Key Laboratory of Microscale Magnetic Resonance, and School of Physical Sciences, University of Science and Technology of China, Hefei 230026, China}
\affiliation{Anhui Province Key Laboratory of Scientific Instrument Development and Application, University of Science and Technology of China, Hefei 230026, China}
\affiliation{CAS Center for Excellence in Quantum Information and Quantum Physics, University of Science and Technology of China, Hefei 230026, China}
\affiliation{Hefei National Laboratory, University of Science and Technology of China, Hefei 230088, China}

\begin{abstract}

The Sb$^{3+}$ ion has garnered significant interest due to its effectiveness in boosting the optical properties of host materials. Among the interesting phenomena is the commonly observed dual-band emission, which has often been interpreted by adopting the phenomenological model that explains the dual-band emission (``ultraviolet band'' and ``visible band'') in Sb-doped $L$PO$_{4}$ ($L$ = Sc, Y, Lu). However, the model for Sb-doped $L$PO$_{4}$ series itself has not been well understood theoretically. In this work, we employ first-principles calculations combined with group-theory analysis to clarify the underlying physical mechanism behind dual-band emission in Sb-doped $L$PO$_{4}$ series. We demonstrate that the dual-band arises from two excited-state equilibrium structures, one exhibits a relatively small distortion with respect to the ground-state equilibrium structure, while the other displays a significantly larger distortion, characteristic of an ``off-center'' configuration. The deviations from the ground-state configuration are dominated by two distinct vibrational modes, $b_2$ and $e$ modes, involving the Jahn-Teller effect and the pseudo Jahn-Teller effect, respectively. Furthermore, charge transition levels and energy barriers calculated using the climbing image nudged elastic band (CI-NEB) method have aided in understanding the relaxation between the two excited-state configurations and the property changes across the Sc, Y, and Lu series. These insights provide a basis for understanding the exotic properties of Sb$^{3+}$ in other hosts and may facilitate the design of optical materials in a broader range of systems involving Sb$^{3+}$ ions.

\end{abstract}

\maketitle

%\tableofcontents

\section{INTRODUCTION}

Sb$^{3+}$ ions have recently garnered widespread attention due to their ability to passivate deep-level defects and enhance carrier mobility within the perovskite layer~\cite{2023NatEnergy-ATO,2024AM-Sn}, broadband emission, anti-thermal quenching, and tunable luminescence properties~\cite{2024jacs-antiTQ,2020jpcl-Sb-NaIn}. By tuning the coordination environment of the doping centers and altering the type of monovalent cations, a broad spectrum of fluorescence colors can be achieved, ranging from blue to near-infrared~\cite{2020jpcl-Sb-NaIn,2019CM-organic-Sb,2020CM-CsInCl5,Sb-pair-organic,2021AFM-Zn}. 

Recently, it has been widely recognized that Sb dopants can produce not only single-band~\cite{2022CM-xia,2020jpcl-Sb-NaIn,1985-CsNaLaCl,2020IC-RbIn,2020CM-Sb-KIn,2021Nano-Cs2CdCl4,1986-Sb-Cs2NaMCl6} but also dual-band~\cite{2019jpcl-xia,2021jacs-Sb-CZC,Sb-Hf-2021CC,2022IC-zou,2023AOM-Sb-CsCdCl3,2022Ceramics,2021AOM-Sb-Xia,Nag-2021jpcl,Gong2022,Zhao2024} emissions, in addition to the host luminescence. For example, in the case of isovalent substitution, Sb$^{3+}$: Cs$_2$NaInCl$_{6}$~\cite{2020jpcl-Sb-NaIn,2020CM-Sb-KIn}, Cs$_2$NaYCl$_{6}$~\cite{1986-Sb-Cs2NaMCl6}, and Rb$_3$InCl$_6$~\cite{2020IC-RbIn} exhibit single-band emission. In contrast, in the case of aliovalent substitution, ${\rm Sb^{3+}: Cs_2(Sn, Zr, Hf)Cl_{6}}$~\cite{2019jpcl-xia,2021jacs-Sb-CZC,Sb-Hf-2021CC} and RbCdCl$_3$, CsCdCl$_3$~\cite{2022IC-zou,2023AOM-Sb-CsCdCl3,2022Ceramics} display dual-band emission. The interpretation of the mechanism behind dual-band emission goes beyond the phenomenological model of isolated ions and includes the following two regimes:
\begin{enumerate}[label=(\roman*)]
 \item Different local minima of the adiabatic potential energy surface (APES) of the excited states~\cite{1970Fukuda-prb}
 \item Distinct luminescent centers, with variations in diverse site occupations and the influence of neighboring point defects accompanying the dopant. 
\end{enumerate} 
The first regime often involves the coupling of orbitally degenerate excited states with degenerate vibrational modes at a high-symmetry site, belonging to the Jahn-Teller effect. 
The second regime often involves either different potential sites in the lattice for occupation, or non-equivalent charges between dopants and the replaced host ions. The presence or absence of an intrinsic defect near the dopant notably affects the material's optical properties. In practice, due to the broad-band nature of the emissions, not all emissions can be resolved, and some sites or local minima may not produce emissions due to quenching. This often allows different phenomenological interpretations of experimental phenomena, thereby posing challenges in determining the true luminescence mechanisms.

More recently, we studied several systems where Sb$^{3+}$ substitutes host ions with higher~\cite{lmz-2021jpcc,2022prb-lmz} or a lower valence state~\cite{2024ic-hrj}, and discovered that intrinsic defects have multiple impacts on aliovalent doping centers. These include the formation of multiple luminescent centers~\cite{lmz-2021jpcc,2022prb-lmz,2023jpcc-hrj,2024ic-hrj} and the stabilization of two or more excited-state equilibrium configurations\cite{2024ic-hrj}.

Actually,  Sb$^{3+}$-related dual-band emission has been reported for a long time in $L\mathrm{PO_{4}}$ ($L$ = Sc, Y, Lu) systems~\cite{1984cpl-YPO4,YPO4}. The intensity ratio of the ``UV band'' (ultraviolet) to the ``visible band'' varies non-monotonically with temperature. 
This phenomenon is particularly intriguing because it lacks the high orbital-degeneracy similar to that described in \cite{1970Fukuda-prb}. Additionally, it 
does not involve aliovalent doping centers, as Sb$^{3+}$ substitutes a single $L^{3+}$ site with D$_{2d}$ point-group symmetry, maintaining the same valence state. Phenomenologically, G.\ Blasse and co-workers~\cite{1984cpl-YPO4,YPO4} still attributed the dual-band emission of Sb$^{3+}$ in the $L$PO$_{4}$ to a sort of Jahn-Teller effect. This  phenomenological model has been applied~\cite{2021AOM-Sb-Xia,Nag-2021jpcl,Gong2022,Zhao2024} to interpret various observed dual-band emission. However, there is still a lack of details in the mechanism and no reliable first-principles calculations to confirm the interpretation of the experimental phenomena.

In this work, we combine group-theoretical analysis of electron-vibration mode coupling with first-principles calculations based on (hybrid) density functional theory to obtain the excited state equilibrium structural characteristics and APES of Sb$^{3+}$-doped rare-earth phosphates. At the D$_{2d}$ ground state equilibrium structure, the three-fold degenerate orbital part of the spin-triplet excited state of isolated Sb$^{3+}$ dopant splits into a non-degenerate B$_2$ level and a doubly-degenerate E level. Depending on the dominance of different electron-vibration couplings, two distinct mechanisms can lead to structural distortions in the excited states. The first mechanism primarily involves linear coupling within the degenerate E level with non-degenerate vibrational modes. This coupling results in distorted equilibrium configurations along the non-degenerate vibration  modes and lifts the degeneracy of the excited level, known as the JT effect. The second mechanism primarily involves coupling between the E level and the non-degenerate B$_2$ level via the $e$ vibrational mode. Additionally, there is second-order coupling within the E level by the $e$-vibration mode, leading to different distorted equilibrium configurations in the plane of the doubly-degenerate $e$ mode, which also lifts the degeneracy, known as  the pseudo-JT effect)~\cite{2013CR-PJTE,2021CR-Bersuker}. These two mechanisms are shown to be responsible for the observed ``UV band'' and ``visible band.

This study provides a in-depth understanding of the dual-band emission of Sb$^{3+}$-doped materials. It also shows that under certain conditions, the interplay of ligand fields, lattice vibrations, and spin-orbit interactions may lead to rich phenomena in excited state dynamics.

\section{METHODOLOGY}

\subsection{Lattice optimization}

The first-principles calculations were carried out with the projector augmented wave method (PAW)~\cite{PAW} implemented in the Vienna ab initio simulation package (VASP) code~\cite{vasp1,vasp2}. The primitive cell of $L\mathrm{PO_{4}}$ ($L$ = Sc, Y, Lu) were optimized with the Perdew-Burke-Ernzerhof revised for solids (PBEsol)~\cite{PBEsol} exchange correlation functional under the 520 eV cutoff energy. The $k$-grids generated by the VASPKIT~\cite{vaspkit} were set as 7 $\times$ 7 $\times$ 7. For comparison, we employed the PBE027 hybrid functional to optimize the lattice constants of $L\mathrm{PO_{4}}$, with a plane-wave cutoff energy set to 520 eV. The exchange fraction $\alpha$ was consistent with our previous studies~\cite{2021ic-fzy}, adopting $\alpha = 0.27$ for three hosts, labeled as PBE027. To reduce computational cost, a 3 $\times$ 3 $\times$ 3 $k$-grids were utilized. The calculated results are summarized in Table~\ref{lattice}. We observed that the lattice parameters obtained using the PBEsol functional agree  better with experimental values, whereas the PBE027 functional tends to systematically underestimate them. Consequently, we adopted the PBEsol-calculated lattice parameters for all subsequent calculations.

\begin{table}[htbp]
\caption{\label{lattice} The lattice constants (in units of \AA) of the tetragonal zircon structure $L$PO$_{4}$ ($L$ = Sc, Y, Lu) conventional cell optimized by PBEsol and PBE027. Relative error of lattice parameter ($\Delta $) given with respect to experimental values.  Considering the computational cost, $k$-grids of PBE027 were set as 3 $\times$ 3 $\times$ 3.  }
  \begin{ruledtabular}
  \begin{tabular}{ccccccccccccc}
  \specialrule{0em}{2pt}{2pt}
  \quad & \quad & PBEsol  & $\Delta $ & PBE027  & $\Delta $& Exp. \cite{LPO4} \\
  \specialrule{0em}{2pt}{2pt}
  \hline
  \specialrule{0em}{2pt}{2pt}
  ScPO$_{4}$& \textbf{a} & 6.583    & 0.14\%  & 6.531  & $-0.65\%$   & 6.574   \\
  \quad     & \textbf{c} & 5.786    & $-0.08\%$  & 5.754  & $-0.64\%$   & 5.791   \\
  \specialrule{0em}{2pt}{2pt}
  YPO$_{4}$ & \textbf{a} & 6.8820   & 0.004\% & 6.839 & $-0.62\%$  & 6.8817    \\
  \quad     & \textbf{c} & 6.0111   & $-0.11\%$ & 5.973 & $-0.74\%$  & 6.0177    \\
  \specialrule{0em}{2pt}{2pt}
  LuPO$_{4}$ & \textbf{a} & 6.744    & $-0.71\%$ & 6.698 & $-1.38\%$  & 6.792    \\
  \quad     & \textbf{c} & 5.903    & $-0.86\%$ & 5.869 & $-1.43\%$  & 5.954    \\              
  \specialrule{0em}{2pt}{2pt}
  \end{tabular}
  \end{ruledtabular}
\end{table}

\begin{figure*}[htbp]
 \centering
 \begin{overpic}[width=1.6\columnwidth]{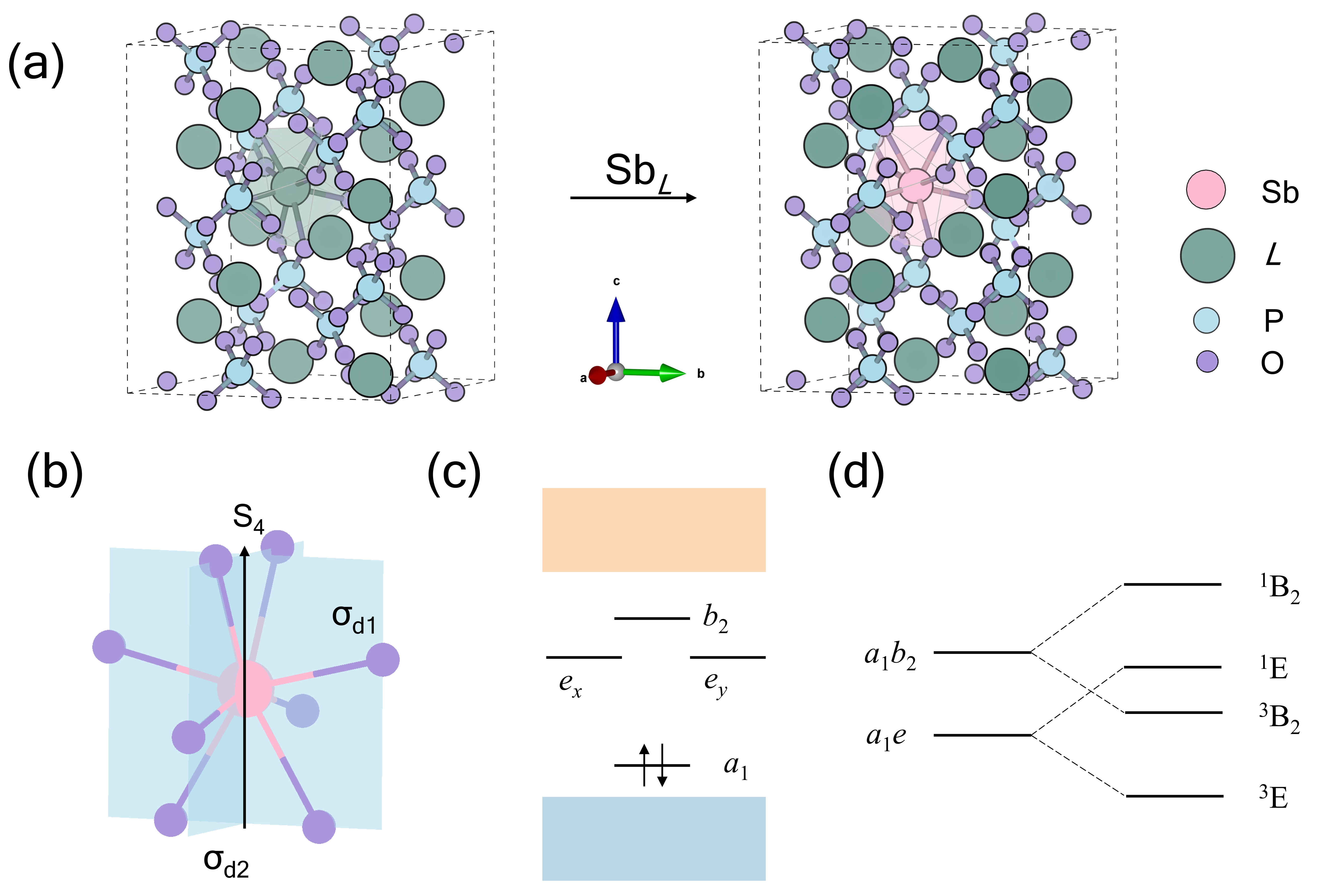}
 \end{overpic} 
 \caption{(a) The left panel shows the crystal structure of $L$PO$_4$ ($L$ = Sc, Y, Lu), and the right panel shows the structure of Sb$^{3+}$: $L$PO$_4$. (b) The local structure of [SbO$_{8}$], with $\sigma_{d1,2}$ representing two diagonal reflection planes. (c) A schematic illustration of the single-particle defect orbital levels within the band gap for the ground-state equilibrium structure, labeled according to the irreducible representations of the D$_{2d}$ point group.
(d) A schematic representation of the splitting of excited-state electron configurations $a_{1}e$ and $a_{1}b_{2}$ into singlet and triplet states due to exchange interaction, without considering spin-orbit coupling.
 }
 \label{Sb_LPO4}
\end{figure*}
\subsection{Electronic structure}

The $k$ paths in the Brillouin zone, going between the high-symmetry $k$ points for the band structure calculations, were generated via the SeeK-path package~\cite{seekpath}. Electronic band structures of the optimized unit cell by PBEsol were calculated using the PBEsol functional and the PBE027 hybrid functional.  A more rigorous approach to compute the accurate bandgap was applying many body perturbation theory within the $GW_{0}$ approximation~\cite{1965GW-Hedin}. We performed partially self-consistent quasiparticle $GW_{0}$ calculations~\cite{1986GW0-Hybertsen}, using the GGA PBEsol functional as the starting point for the $GW_{0}$ approach. The details of the parameters used in the $GW_{0}$ calculation are presented in the Supplemental Material~\cite{supplemental_material}.

\subsection{Excited-state calculations}

Supercells with new base vectors ($\textbf{a} - \textbf{b}, \textbf{a} + \textbf{b}, 2\textbf{c}$) containing 96 atoms and the $\Gamma$-point only in the Brillouin zone were employed to study the defect-related properties. The Sb atom replaces an $L$ atom at a site closest to the center of the supercell. The energy cutoff was set to 520 eV, and the single $\Gamma$-point was applied to sample the Brillouin zone of the supercells for calculations of hybrid functionals. The excited state equilibrium structures were optimized by PBEsol and PBE027 to compare the differences of these functionals. Calculations of dopant-related lowest excited states were performed by setting spin quantum numbers to a triplet state when spin-orbit coupling (SOC) was turned off, and by the constrained occupation method when SOC was included. To demonstrate that the effect of finite size can be ignored, we use Sb$^{3+}$-doped YPO$_4$ as an example and compare the results of supercells of two different sizes, as shown in Table S2 in the Supplemental Material~\cite{supplemental_material}. 

To determine the energy barrier $\delta $ between the excited-state equilibrium structures, we utilized the climbing image nudged elastic band (CI-NEB) method~\cite{2000JCP-Henkelman1,2000JCP-Henkelman2}. This method constructs the minimum energy path in atomic-configuration space by interpolating a series of interconnected images between the equilibrium structures.

\subsection{Charge transition levels}

The formation energy of a defect $D$ in the charge state $q$ is defined by~\cite{2014rmp-cc}
\begin{equation}
 \begin{split}
 \label{formation}
 E^{f}[D^q ,E_{\rm {F}}] &= [E_{\rm{tot}}[D^{q}] + E_{\rm {corr}}] - E_{\rm{tot}}[\rm{bulk}] \\
 &- \sum_{i}n_{i}\mu_{i} + qE_{\rm F},
\end{split}
\end{equation}
where $E_{\rm{tot}}[D^{q}]$ and $E_{\rm{tot}}[\mathrm{bulk}]$ are the calculated total energies of a supercell with or without defect $D$ with charge $q$; $n_{i}$ represents the change in number of atom $i$ due to defect $D$, which is added to ($n_{i} > 0$) or removed from ($n_{i} < 0$) the pristine supercell; $\mu_{i}$ is the chemical potential for atom $i$ constrained by the coexistence of related substances, and $E_{\rm F}$ is the electron Fermi energy, which is determined by overall charge neutrality of all charged defects. The term $E_{\rm {corr}}$ is to correct the error in the supercell method under the periodic boundary condition in a practical DFT calculation. It usually includes two parts: the electrostatic image charge interaction and the potential alignment between the pristine supercells and the charged defect. The corrections that extend the scheme of Frysoldt, Neugebauer, and Van de Walle to an anisotropic medium (eFNV) are applied using the code developed by Kumagai~\cite{2009FNV, 2014eFNV,pydefect}. 

The thermodynamic charge transition level (CTL) $\epsilon (q_{1}/q_{2})$ is defined as the value of Fermi energy where the formation energies of charge states $q_{1}$ and $q_{2}$ are equal. From Eq.~\ref{formation}, we get 

\begin{equation}
 \label{ctl}
 \epsilon (q_{1}/q_{2}) = \frac{E^{f}[D^{q_{2}}, 0] - E^{f}[D^{q_{1}}, 0]}{q_{1} - q_{2}}.
\end{equation}
In principle, the defect changes from the higher to the lower valences as the Fermi level increases across the charge transition level. When $\epsilon (q_{1}/q_{2})$ ($q_1=q_2+1$) falls in the forbidden band, the gaps between $\epsilon (q_{1}/q_{2})$ and band edges are the energies released when an electron and a hole are captured by $D^{q_1}$ and $D^{q_2}$, respectively. To compare the positions of transition levels in different host materials, we selected the vacuum level as the potential energy zero point. The details of calculationn of  vacuum-referred binding energy (VRBE) are presented in the Supplemental Material~\cite{supplemental_material}.

\subsection{Configuration coordinate diagrams}

The interpolated structure $\mathbf{R}_{\alpha}$ between the initial equilibrium structure $\mathbf{R}^{i}_{\alpha}$ and the final equilibrium structure $\mathbf{R}^{f}_{\alpha}$ is obtained via linear interpolation. In the one-dimensional model adopted here, a signed generalized configuration coordinate $Q$ is introduced for every interpolated structure, following Ref.~\cite{2014rmp-cc,2014prb-cc}:
\begin{equation}
    \label{cc}
     Q^{2} = \sum_{\alpha} m_{\alpha} (\mathbf{R}_{\alpha} - \mathbf{R}^{i}_{\alpha })^{2},
\end{equation}
where $m_{\alpha}$ is the mass of atom $\alpha$. Note that $Q=0$ when $\mathbf{R}_{\alpha}$ corresponds to $\mathbf{R}^{i}_{\alpha}$.  

We employed the $\Delta$SCF~\cite{SCF} method, which involves determining the excited-state equilibrium structure and computing the total energy of the (lowest) excited state relative to the ground state. Consequently, the configuration coordinate diagram was constructed and the energies of vertical transitions were obtained. We also examined the impact of spin-orbit coupling on excited-state dynamics, as antimony is a fourth period-post-transition metal. The energy and force tolerances were set to 10$^{-5}$ eV and $0.01~$eV/\r{A} per atom, respectively. 

\section{RESULTS AND DISCUSSION}

\subsection{Hosts' geometric and electronic structures}

$L\mathrm{PO_{4}}$ ($L$ = Sc, Y, Lu) belong to the tetragonal systems with I$4_{1}/$amd space group (No.\ 141)~\cite{LPO4}, the crystal structure is shown as Figure~\ref{Sb_LPO4}(a). The metal ions $L^{3+}$ are coordinated by eight oxygen atoms forming a dodecahedron with site symmetry D$_{2d}$. The [$L$O$_{8}$]$^{15-}$ polyhedra are connected by [PO$_{4}$]$^{3-}$ tetrahedra, and the [$L$O$_{8}$]$^{15-}$ polyhedra share edges with each other. Geometrically optimized lattice constants of $L\mathrm{PO_{4}}$ with the PBEsol functional are in good agreement with the experimental values~\cite{LPO4} and with previous calculated values for $\mathrm{YPO_{4}}$ and LuPO$_{4}$~\cite{2021ic-fzy}, as shown in Table~\ref{lattice}. The band gap of the hosts are calculated using PBEsol, PBE027, and $GW_{0}$ methods. The results reveal a significant underestimation of the band gap by PBEsol compared to experimental values and other computational approaches. In contrast, the band gap calculated using PBE027 is in good agreement with both the $GW_{0}$ results and experimental measurements, thereby validating the appropriateness of our chosen $\alpha$ parameter. The band structure of $L\mathrm{PO_{4}}$ obtained with $GW_0$ is shown in Figure S1~\cite{supplemental_material}. The electronic properties of the $L\mathrm{PO_{4}}$ series are all similar. The valence bands are composed of O--p and the conduction bands are all composed of d orbitals for $\mathrm{ScPO_{4}}$, $\mathrm{YPO_{4}}$, and LuPO$_{4}$. Further details of the results, along with the experimental values, are provided in Table S1 \cite{supplemental_material}.

\begin{figure*}[htbp]
 \centering
 \begin{overpic}[width=2.0\columnwidth]{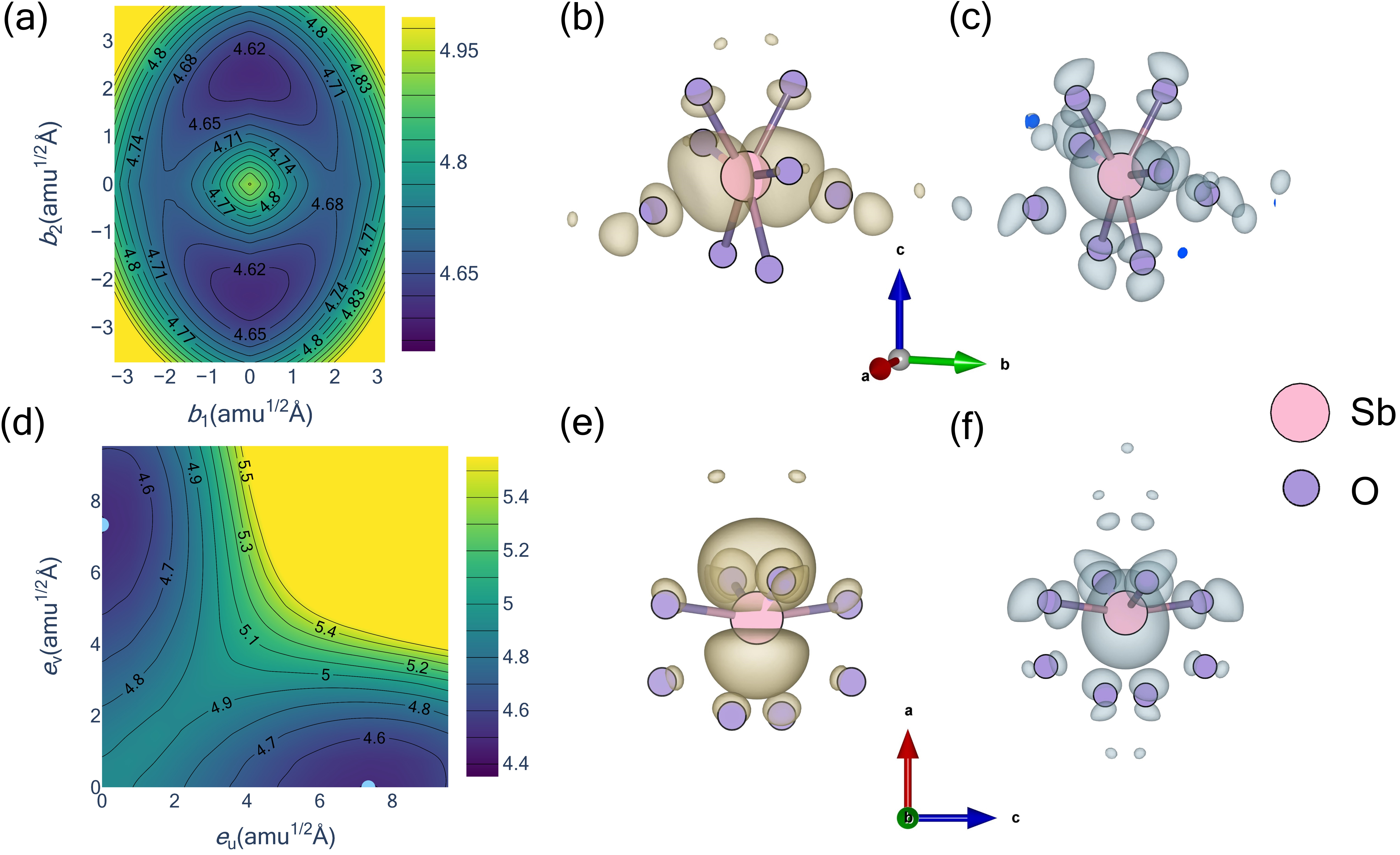}
 \end{overpic} 
 \caption{(a), (d) The excited-state APES of Sb$^{3+}$ in $\mathrm{YPO_{4}}$ along the equivalent $b_{1,2}$ and $e_{u,v}$ modes, respectively, were obtained by the PBEsol functional. The calculations were performed by $\Delta$SCF with setting spin quantum numbers without including SOC. The excited-state equilibrium structure was obtained by applying an initial distortion to the ground-state equilibrium structure, according to the vibrational modes described in the Appendix B. The squared orbital functions of the electron orbital (b), (e) and hole orbital (c), (f) were calculated at the equilibrium excited-state structures dominated by the distortion of the equivalent $b_{2}$ and $e$ vibrational modes.}
 \label{b_e_all}
\end{figure*}

\subsection{Group-theoretical analysis}

In the ground state, the local environment of Sb$^{3+}$ in $L\mathrm{PO_{4}}$ exhibits D$_{2d}$ symmetry. Upon excitation of Sb$^{3+}$, it rapidly undergoes non-radiative relaxation to approach the local equilibrium structures of the lowest electronic excited state ($a_{1}e$) $^{3}$E in Fig.~\ref{Sb_LPO4}(c,d). We consider the following two regimes of structure relaxations that lower the degeneracy and energy of $^{3}$E: (i) the Jahn-Teller effect, denoted as $\mathrm{E} \otimes (b_1 + b_2)$, which reduces the symmetry to C$_{2v}$, and (ii) the pseudo Jahn-Teller effect~\cite{2021CR-Bersuker}, represented by $(\mathrm{B}_{2} + \mathrm{E}) \otimes e$, which lowers the symmetry to C$_{2}$.
The capital letters represent the irreducible representations of electronic states, while lowercase letters denote the irreducible representations of  vibrational modes.

It should be noted that the electron states couple to all vibrational modes at the same time. For simplicity and for a clearer picture, we have considered two independent cases governed by  JT and pesudo-JT effects, where distortions are dominated by different vibration modes.

The dependence of the electron-nuclear interaction $V(r, Q)$ on $Q$ can be expanded as~\cite{2006Bersuker},
\begin{equation}
 \begin{split}
 W (r,Q) &= V(r,Q) - V(r,Q_{0}) \\
 &= \sum_{\alpha} \dfrac{\partial V}{\partial Q_{\alpha}} Q_{\alpha} + \dfrac{1}{2} \sum_{\alpha \beta} \dfrac{\partial^{2} V}{\partial Q_{\alpha} \partial Q_{\beta}} Q_{\alpha} Q_{\beta} + \cdots
 \end{split}
 \label{vib_interaction}
\end{equation}
The matrix elements of the expansion coefficients in Eq.~\eqref{vib_interaction} are called vibronic constants~\cite{2006Bersuker}, which govern the changes in the electronic structure influenced by nuclear displacements or phonon modes.

For the $\textrm{E} \otimes (b_1 + b_2)$ regime, the linear electron-phonon coupling term combined with the parabolic vibrational potential determines the extrema points in the APES. Referring to Table~\ref{CG1} in Appendix A, the matrix representation of the linear term in Eq.(\ref{vib_interaction}) is: 
\begin{equation}
 W = 
 \begin{bmatrix}
 F_{2}Q_{2} & F_{1}Q_{1} \\
 F_{1}Q_{1} & -F_{2}Q_{2} \\
 \end{bmatrix},
 \label{b1_b2}
\end{equation}
with 
\begin{eqnarray} 
F_{1} &=& \left \langle v \left |\left (\dfrac{\partial V}{\partial Q_{b_{1}} }\right )_{0} \right |u \right \rangle , \\F_{2} &=& \left \langle u\left |\left (\dfrac{\partial V}{\partial Q_{b_{2}} }\right )_{0} \right |u \right \rangle  = - \left \langle v \left |\left (\dfrac{\partial V}{\partial Q_{b_{2}} }\right )_{0} \right |v \right \rangle . 
\end{eqnarray}
Here, $|u \rangle  $ and $|v \rangle $ are the orbital wave functions of the two E states, which transform in the same way as the $x$ and $y$ components of a $p$ orbital. The solution of the combination of Eq.~\eqref{b1_b2} with the parabolic vibrational potential produces two pairs of equilibrium structures, one pair along the $Q_{b_{1}}$ direction and the other along the $Q_{b_{2}}$ direction.

For the regime ($^{3}$B$_{2}$ + $^{3}$E) $\otimes$ $e$, a Mexican hat-like potential energy surface is produced if only the linear vibronic constants are considered. By including the quadratic terms~\cite{2006Bersuker}, the quantity defined by Eq.\ (\ref{vib_interaction}) can be written as the following matrix:
\begin{equation}
 \begin{bmatrix}
 - \Delta + G_1 (Q_{u}^{2} - Q_{v}^{2}) & G_2 Q_{u} Q_{v} & F Q_{u} \\
 G_2 Q_{u} Q_{v} & - \Delta - G_1 (Q_{u}^{2} - Q_{v}^{2}) & - F Q_{v} \\
 F Q_{u} & - F Q_{v} & \Delta
 \end{bmatrix},
 \label{e}
\end{equation}
where 2$\Delta$ denotes the energy gap between the $^{3}$B$_{2}$ (with the orbital wave function $|b_2\rangle$) and $^{3}$E, $F$ denotes the pseudo Jahn-Teller linear vibronic constant expressed as
\begin{equation}
 F = \left \langle b_2 \left |\dfrac{\partial V}{\partial Q_{u} } \right |u \right \rangle  = - \left \langle b_2\left|\dfrac{\partial V}{\partial Q_{v} } \right |v \right \rangle ,   
\end{equation}
and $G_1$ and $G_2$ are the pseudo Jahn-Teller quadratic vibronic constants given  as
\begin{eqnarray}
2G_1 &=& \left \langle u \left | \dfrac{\partial^{2} V}{\partial Q_{u}^{2}} \right | u \right \rangle  = - \left \langle v \left | \dfrac{\partial^{2} V}{\partial Q_{v}^{2}} \right | v \right \rangle  , \\
2G_2 &=& \left \langle u \left | \dfrac{\partial^{2} V}{\partial Q_{u} \partial Q_{v} } \right | v \right \rangle  = \left \langle v \left | \dfrac{\partial^{2} V}{\partial Q_{u} \partial Q_{v}} \right | u \right \rangle 
\end{eqnarray}
In polar coordinates for the degenerate vibration modes, $Q_{u} = \rho \ \rm{cos\phi}$ and $Q_{v} = \rho \ \rm{sin\phi}$. The extrema points are at $\phi = n\pi/4$ ($n = 0, 1, 2, \cdots, 7$). If $G_{1}$ $>$ $G_{2}$, minima occur at $\phi = n\pi/4$ with even $n$, while odd $n$ values form saddle points. Conversely, if $G_{1}$ $<$ $G_{2}$, odd $n$ values correspond to minima, while even $n$ correspond to saddle points (see Chapter 4 of Ref~\cite{2006Bersuker}).

\begin{table*}[htbp]
 \caption{\label{table4} Excitation (exc.), emission (emi.), and their zero-phonon line (zpl) energies  for Sb$^{3+}$ in $L\mathrm{PO_{4}}$ ($L=\textrm{Sc, Y, and Lu}$) calculated using PBEsol and PBE027 functions, together with experimental results (in units of eV). Here, JT and pJT denote the equilibrium structures of the excited states associated with the Jahn-Teller and pseudo Jahn-Teller effects, respectively, while the JT* here represents the equilibrium structure obtained by applying a small distortion to the JT equilibrium structure towards the pJT equilibrium structure. }
 \begin{ruledtabular}
 \begin{tabular}{ccccccccccccccccccccccccccc}
  \specialrule{0em}{2pt}{2pt}
 \quad & $L\mathrm{PO_{4}}$ & \multicolumn{7}{c}{ PBEsol} & & \multicolumn{7}{c}{ PBE027} & \multicolumn{2}{c}{Exp.~\cite{1984cpl-YPO4,YPO4}} \\  \specialrule{0em}{1pt}{1pt} \cline{3-9} \cline{11-17} 
\specialrule{0em}{2pt}{2pt}
\quad & \quad & \quad & \multicolumn{2}{c}{JT} & \multicolumn{2}{c}{JT*} & \multicolumn{2}{c}{pJT} & & \quad &\multicolumn{2}{c}{JT} & \multicolumn{2}{c}{JT*} & \multicolumn{2}{c}{pJT} & \\
\specialrule{0em}{2pt}{2pt}
 \quad &  \quad & exc. & zpl & emi. & zpl & emi. & zpl & emi. & \quad & exc.& zpl & emi. & zpl & emi. & zpl & emi. & exc. & emi.\\
\specialrule{0em}{2pt}{2pt}
\hline 
\specialrule{0em}{2pt}{2pt}
with  & Sc &\multicolumn{7}{c}{/}& & 5.02\footnote{Estimated by the energy difference of $ns\rightarrow np$ single-electron excitation calculated at the equilibrium structure of the ground state.} & 4.79 & 4.42 &\multicolumn{2}{c}{/} & 4.63 & 2.91 & 5.17 & 4.13, 3.40 \\
\specialrule{0em}{2pt}{2pt}
SOC & Y & 4.63 & 4.51 & 4.21 & 4.47 & 3.98 & 4.45 & 3.15 & & 4.94 & 4.67 & 4.20 & 4.59 & 3.89 & 4.44 & 2.43 & 5.06 & 4.28, 2.99 \\ 
\specialrule{0em}{2pt}{2pt}
\quad & Lu & 4.72 & 4.63 & 4.39 & 4.58 & 4.08 & 4.55 & 3.40 & & 5.02 & 4.80 & 4.37 & 4.71 & 4.03 & 4.61 & 2.72 & 5.06 & 4.28, 3.10 \\ 
\specialrule{0em}{2pt}{2pt}
\hline 
\specialrule{0em}{2pt}{2pt}
without & Sc &\multicolumn{7}{c}{/}& & 5.32$^{\text{a}}$ & \multicolumn{4}{c}{/} & 4.66 & 2.89 & 5.17 &4.13, 3.40  \\
  \specialrule{0em}{2pt}{2pt}
SOC & Y & 4.91 & 4.60   & 4.23  & 4.55 & 3.96  & 4.50 & 3.10  & & 5.25 & 4.74  & 4.26 & 4.65 & 3.90 &  4.48 & 2.44 & 5.06 & 4.28, 2.99  \\ 
 \specialrule{0em}{2pt}{2pt}
\quad & Lu & 5.02 & 4.74 & 4.42 & 4.66 & 4.07  & 4.61 & 3.33  & & 5.33 & 4.88  & 4.42 & 4.77 & 4.04 & 4.64 & 2.74 & 5.06 & 4.28, 3.10  \\   
 \specialrule{0em}{2pt}{2pt}
 \end{tabular}
 \end{ruledtabular}
\end{table*}

\subsection{Verification via First-Principles Calculations}

Using the model Hamiltonian for Jahn-Teller and pseudo Jahn-Teller effects, we have pinpointed the vibrational modes that significantly impact the potential energy surface. Interestingly, in the vibrational modes of the local structure [SbO$_8$], the irreducible representations of the displacements of the central ion match those of the vibration modes involved in the Jahn-Teller ($b_2$) and pseudo Jahn-Teller effects ($e$), as shown in Appendix~\ref{vib_mode}. Consequently, by starting from geometric structures with different initial displacements of the central ion over the equilibrium geometric structure of the ground state, we obtained different (constrained) excited-state equilibrium geometric structures, i.e., minima on the APES, dominated by displacements of different equivalent modes. It is important to note, however, that in real systems, there may be multiple vibrational modes that correspond to the same irreducible representation, as shown in Appendix~\ref{vib_mode}. Therefore, the aforementioned analysis of vibration modes can be understood as being based on effective vibration modes composed of many modes of the irreducible representation of the point group.

The $L\mathrm{PO_{4}}$ series are isostructure and contain only one site for $L$. Hence, we take Sb$^{3+}$ in $\mathrm{YPO_{4}}$ as an example to demonstrate how the two excited-state local minima dominated by the Jahn-Teller and the pseudo Jahn-Teller effects individually affect the potential energy surface. 

Consistent with the analysis of the regime involving the Jahn-Teller effect, as illustrated in Figure~\ref{b_e_all} (a), the results of first-principles calculations reveal the presence of extrema on the APES along the equivalent $b_{1}$ and $b_{2}$ modes. Specifically, there is a pair of equivalent minima along the $b_{2}$ mode, where the symmetry of the system decreases from D$_{2d}$ to C$_{2v}$, while a pair of equivalent saddle points appears along the $b_{1}$ axis. The squared orbital functions of the electron (with component Sb-5p) and hole (with component Sb-5s) at the equilibrium excited-state structure dominated by the equivalent $b_{2}$ mode are shown in Figure~\ref{b_e_all} (b, c). In Figure~\ref{b_e_all} (d), we present the APES resulting from the pseudo Jahn-Teller effect, and the squared orbital functions of the electron (Sb-5p) and hole (Sb-5s) at the equilibrium excited-state structure dominated by the equivalent $e$ mode are shown in Figure~\ref{b_e_all} (e, f). Following the analysis of the regime involving the pseudo Jahn-Teller effect, we obtained four equivalent excited-state structures with the same minimal energy, which correspond to slight structural distortions of the central ion along the (+x, +y), (+x, -y), (-x, +y), and (-x, -y) directions, where the symmetry decreases from D$_{2d}$ to C$_{2}$. To simplify the computation, we selected two of these structures to construct the APES. The calculation results are in agreement with our expectations. In general, the APES dominated by the Jahn-Teller effect and the pseudo Jahn-Teller effect exhibit significant differences. According to the characteristics of the excited-state structures dominated by the two different modes, we term the excited-state configuration dominated by the $b_{2}$ mode as the ``high-symmetry configuration'', labeled as JT, and the one dominated by the $e$ mode as the ``off-center configuration”, labeled as pJT. The vertical transition energies between JT and pJT, as well as the zero-phonon line energies, are summarized in Table~\ref{table4}.

\begin{figure*}[htbp]
 \centering
 \begin{overpic}[width=1.8\columnwidth]{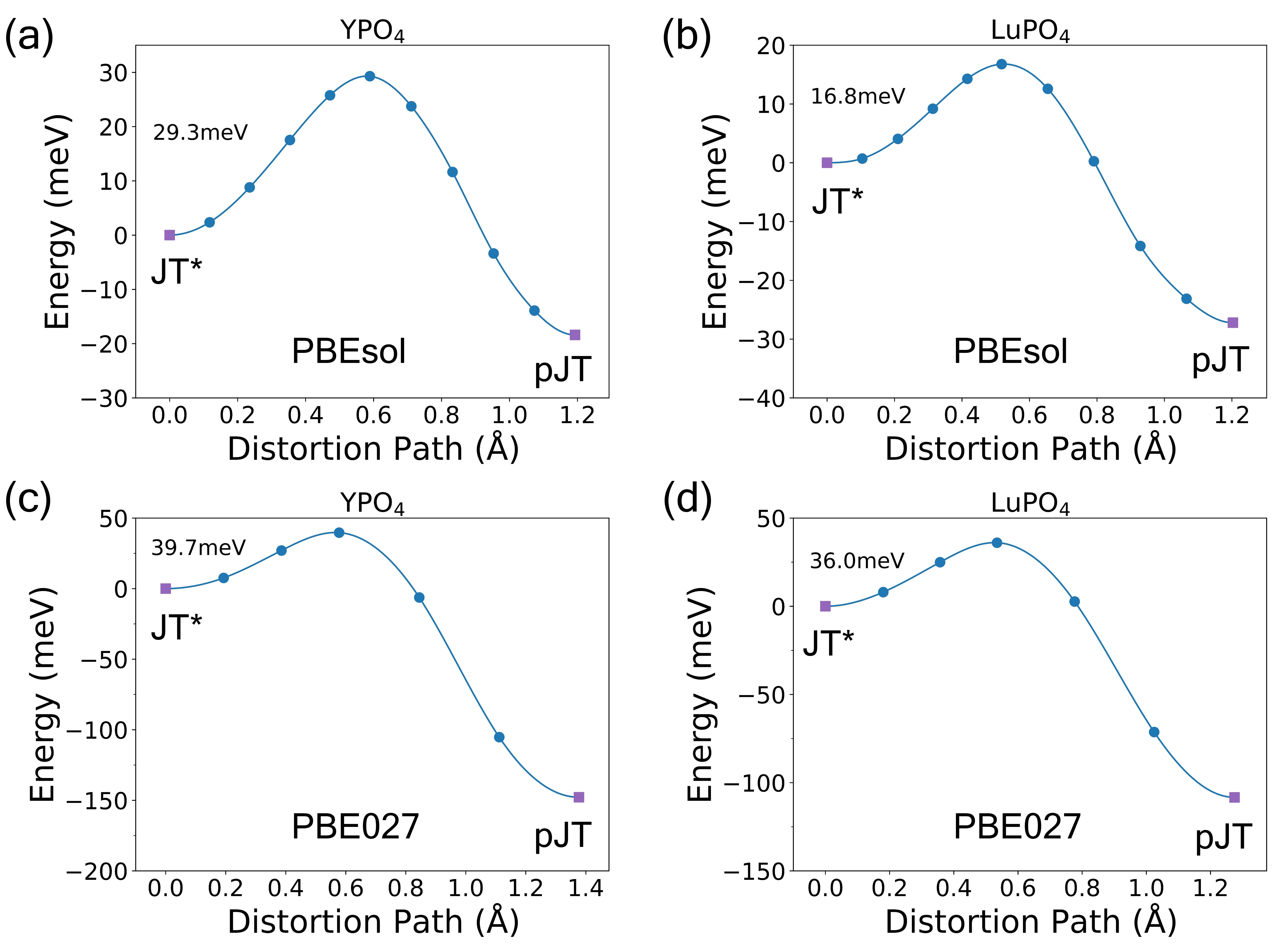} 
 \end{overpic} 
 \caption{
Energy barriers between JT* and the pJT of YPO$_{4}$ (a,c) and LuPO$_{4}$ (b,d) calculated using the climbing image nudged elastic band (CI-NEB) method~\cite{2000JCP-Henkelman1,2000JCP-Henkelman2}, including SOC. For the PBEsol functional, 9 images were inserted between the two equilibrium structures, while for the PBE027 hybrid functional, 5 images were used. 
 }
 \label{b2_e}
\end{figure*}

Although these results align well with the group-theoretical analysis presented in the previous section, it is important to note that couplings to different vibration modes coexist. Therefore, the stability of the obtained structures must be verified against small symmetry-breaking distortions. When ionic relaxation was initiated with these distortions applied to the Jahn-Teller (JT) structure, a new excited-state equilibrium structure with C$_{1}$ symmetry, referred to as JT*, was achieved. As indicated in Table~\ref{diff_bond}, the JT* structure exhibits an average Sb-O bond length that closely resembles that of the JT structure. Furthermore, the root-mean-square deviation of the Sb-O bond lengths in the JT* structure, relative to the ground-state equilibrium structure, is also very close to that of the JT structure. This indicates that the JT* excited-state equilibrium structure remains predominantly governed by the Jahn-Teller effect. Moreover, the pJT structure is stable as it is essentially recovered after ionic relaxation from configurations with small introduced distortions. 

To further verify that the JT* structure is not a saddle point and to quantitatively determine  the energy barrier between JT* and pJT, we employed the CI-NEB method~\cite{2000JCP-Henkelman1,2000JCP-Henkelman2}. For the PBEsol calculations, we inserted 9 images, whereas for the computational more intensive PBE027 calculations, we used 5 images, as shown in Figure~\ref{b2_e}. We find that the energy barrier $\delta$ and zero-phonon line energy difference $\Delta E_{\mathrm{zpl}}$ are functional-dependent. Detailed information on the energy barrier $\delta$ and the zero-phonon line energy difference $\Delta E_{\mathrm{zpl}}$ between JT* and pJT is provided in Table~\ref{barrier}. In the YPO$_4$ system, the barrier calculated using PBEsol is 29.3 meV, while that obtained with PBE027 is 39.7 meV. Both results are in satisfactory agreement with the experimental results~\cite{1984cpl-YPO4,YPO4}. However, the magnitude of $\Delta E_{\mathrm{zpl}}$ differs significantly between the two functionals: PBEsol gives 18 meV, whereas PBE027 gives 148 meV. The PBEsol value deviates substantially from the experimental data, while the PBE027 result shows closer agreement. Moreover, the energy barrier between JT* and pJT obtained without considering SOC is found to be similar to that calculated with SOC included, as shown in Figure S4~\cite{supplemental_material} and Figure~\ref{b2_e}. This indicates that the inclusion of spin-orbit coupling has a relatively minor effect on the energy barrier between these two configurations.

Despite the fact that the PBEsol underestimates the bandgap of $\mathrm{YPO_{4}}$, the calculated vertical transition energies of Sb$^{3+}$ in $\mathrm{YPO_{4}}$ align well with the experimental results~\cite{1984cpl-YPO4,YPO4}, as shown in Table~\ref{table4}. The PBE027 hybrid functional also gives reasonable results, but the calculated emission energies of the pJT are underestimated compared to the experimental values, similar to the case with halide systems~\cite{2023jpcc-hrj,2022prb-lmz,2024ic-hrj}. This suggests that the vertical transition energies calculated at the  excited-state equilibrium structures dominated by the $e$ mode distortion are more sensitive to the choice of the fraction of exact exchange in the hybrid functional, $\alpha$, than those calculated at those equilibrium structures dominated by the $b_{2}$ mode distortion. In the case of polarons, Sio et al.\ indicated that the formation energy of polaron increases with $\alpha$ (see Supplemental Note 1 in~\cite{2022Natphy}). However, for excitons or excitonic polarons~\cite{2024PRL-exciton,2024PRB-exciton}, crystal structure distortions due to electron-hole interactions may not uniformly impact the excited- and ground-state structures. A detailed examination of this scenario extends beyond the focus of this paper and will not be further elaborated upon here.

\begin{table}[]
 \centering
 \caption{\label{diff_bond} Average bond length $\bar{l}$ (\AA) of Sb-O bonds in the [SbO$_8$] local structure for different equilibrium configurations, along with the root-mean-square deviation $\Delta l$ (\AA) of bond lengths between excited-state and ground-state configurations ($\Delta l = \sqrt{\frac{1}{8}\sum_i (l^{\text{ES}}_{i} - l^{\text{GS}}_{i})^2}$), including SOC. }
 \begin{ruledtabular}
 \begin{tabular}{*{14}{c}}
 \specialrule{0em}{2pt}{2pt}
 \quad & $L$PO$_{4}$ & GS &\multicolumn{2}{c}{JT} & \multicolumn{2}{c}{JT*} & \multicolumn{2}{c}{pJT} \\
  \specialrule{0em}{2pt}{2pt}
 \quad & \quad & $\bar{l}$ & $\bar{l}$ & $\Delta l$ &  $\bar{l}$ & $\Delta l$  & $\bar{l}$ & $\Delta l$ \\
 \specialrule{0em}{2pt}{2pt}
 \hline 
 \specialrule{0em}{2pt}{2pt}
 \specialrule{0em}{2pt}{2pt}
 PBEsol & Y & 2.360 & 2.368 & 0.08 & 2.379 & 0.13 & 2.440 & 0.27 \\
 \specialrule{0em}{2pt}{2pt}
 \quad & Lu & 2.332 & 2.334 & 0.07 & 2.348 & 0.12 & 2.385 & 0.22  \\
  \specialrule{0em}{2pt}{2pt}
  \hline 
   \specialrule{0em}{2pt}{2pt}
 PBE027 & Sc & 2.302 & 2.283 & 0.08 & \multicolumn{2}{c}{/} & 2.369 & 0.28   \\
  \specialrule{0em}{2pt}{2pt}
   \quad & Y & 2.362 & 2.362 & 0.10 & 2.375 & 0.16 & 2.491 & 0.41 \\
\specialrule{0em}{2pt}{2pt}
    \quad & Lu & 2.332 & 2.331 & 0.09 & 2.346 & 0.14 & 2.427 & 0.33 \\
\specialrule{0em}{2pt}{2pt}
 \end{tabular}
 \end{ruledtabular}
\end{table}

\subsection{Impact of spin-orbit coupling}

SOC has a profound influence on the energy level structure of an free Sb$^{3+}$ ion~\cite{2021AOM-Sb-Xia}. However, within a crystal lattice, the interplay between SOC and the structural distortions induced by the Jahn-Teller effect tends to suppress the effect of SOC~\cite{1965PRB-Ham,2020PRX-JTE-SOC}. Consequently, the calculated emission energies of Sb$^{3+}$ in $L\mathrm{PO_{4}}$ are not significantly affected by inclusion of SOC, as shown in Table~\ref{table4}. 

Although Bi$^{3+}$ shares a similar electronic configuration with Sb$^{3+}$, the optical properties of Bi: YPO$_{4}$ and LuPO$_{4}$~\cite{2021ic-fzy,2018jmcc-Bi,2017OM-Bi,2015OM-Bi} and their variation with temperature exhibit distinct characteristics compared to those of Sb: YPO$_{4}$ and LuPO$_{4}$. This variation arises from the substantial difference in the strength of SOC, $\xi$. By utilizing virtual Kohn-Sham orbitals, we can estimate the SOC strength~\cite{2021ic-fzy}. For instance, the SOC parameter, $\xi$, is calculated to be 0.47 eV for Sb: YPO$_{4}$, but a much larger value of 1.27 eV is obtained for Bi: YPO$_{4}$. The striking difference in SOC parameters leads to significant alterations in the potential energy surfaces~\cite{2020PRX-JTE-SOC}, which in turn influence the optical properties.

\begin{figure}[]
 \centering
 \begin{overpic}[width=1.0\columnwidth]{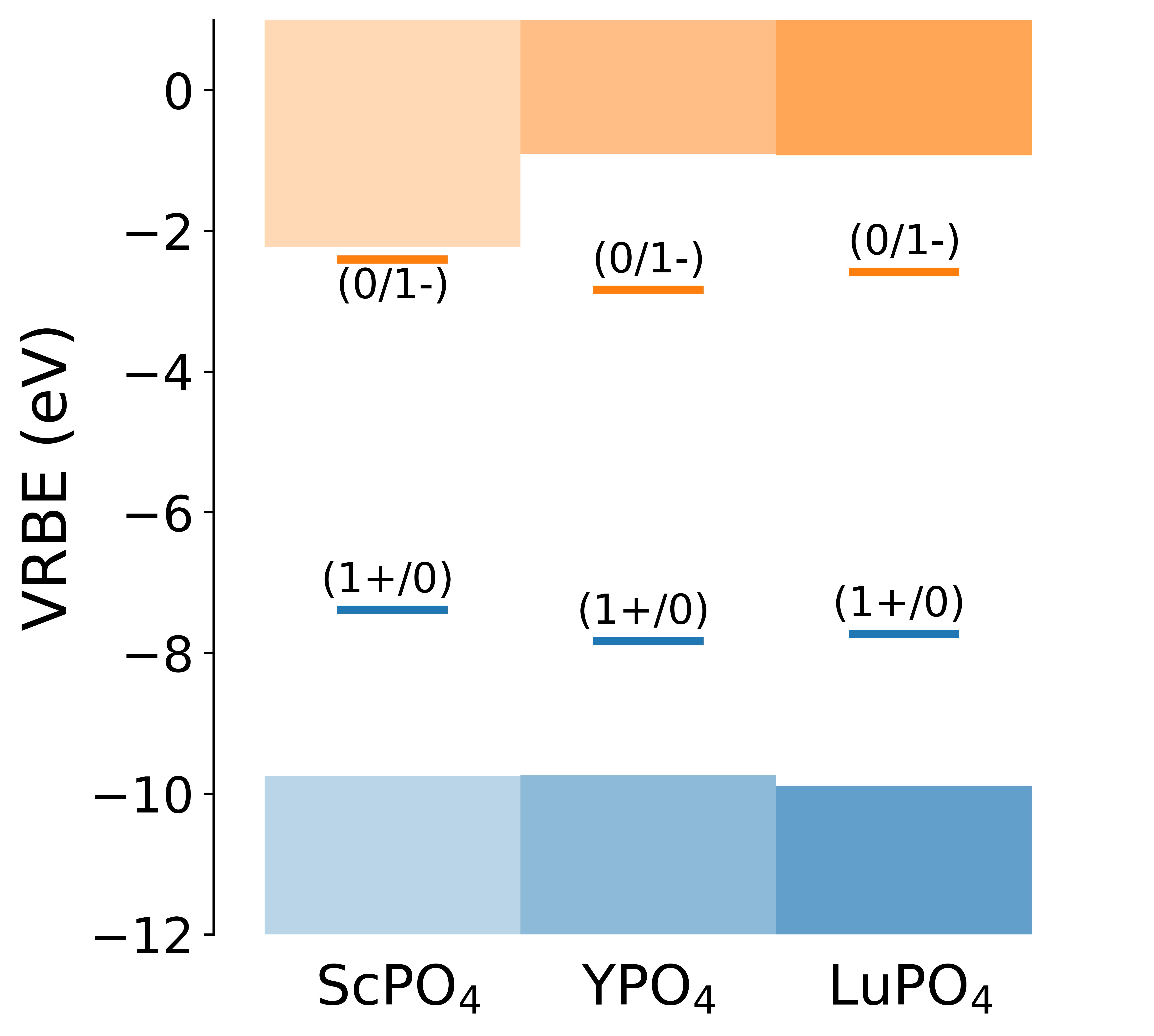}
 \end{overpic} 
 \caption{Vacuum referred binding energy of thermodynamic charge transition levels of Sb$^{3+}$ in $L\mathrm{PO_{4}}$ ($L$ = Sc, Y, Lu) calculated by PBE027.}
 \label{tran_level}
\end{figure}

\subsection{Differences among the $L\mathrm{PO_4}$ series}

To clarify the differences among the three systems, we computed the thermodynamic charge transition levels, as depicted in Figure~\ref{tran_level}. Figure~\ref{tran_level} demonstrates that the thermodynamic charge transition level of (0/1-) in the ScPO$_4$ system is notably closer to the conduction band minimum (CBM) than in the other two systems, indicating a higher probability of Sb$^{3+}$ ionization in ScPO$_4$. Additionally, we attempted to introduce a small distortion to the JT structure of Sb$^{3+}$: ScPO$_{4}$ along the direction of the pJT structure. However, during structural optimization, the electron tended to relax into the conduction band, preventing the attainment of a JT* structure analogous to those observed in the YPO$_4$ and LuPO$_4$ systems. This behavior is attributed to the fact that the transition level of Sb$^{3+}$ (0/1-) lies very close to the conduction band. In contrast, for the Y/LuPO$_{4}$ systems, the thermodynamic charge transition levels of Sb$^{3+}$ are farther from the band edges, resulting in superior luminescent properties compared to ScPO$_{4}$~\cite{YPO4,1984cpl-YPO4}. 

Furthermore, we examined the variations in the average bond length $\bar{l}$ of Sb-O bonds within the [SbO$_8$] local structure for different equilibrium configurations, as well as the root-mean-square deviation $\Delta l$ of bond lengths between excited-state and ground-state configurations, as shown in Table~\ref{diff_bond}. For the JT and JT* structures, the $\Delta l$ values across the three systems $L$PO$_4$ are found to be relatively similar and show little correlation with the host $L^{3+}$ cation. In contrast, for the pJT structure, YPO$_4$ displays the largest deviation from the ground-state equilibrium configuration, regardless of whether calculated using PBEsol or PBE027. 

In the LuPO$_{4}$ system, the energy barrier between JT* and pJT calculated using the CI-NEB method is smaller than that in the YPO$_{4}$ system, as shown in Table~\ref{barrier} and Figure~\ref{b2_e}. Notably, the barrier obtained with PBEsol in the LuPO$_{4}$ system is only about half of that in the YPO$_{4}$ system. This suggests that at low temperatures, the LuPO$_{4}$ system is more likely to relax from the JT* structure to the pJT structure. This finding is consistent with experimental observations, where both the ``UV band'' and ``visible band'' are simultaneously detected at low temperatures~\cite{1984cpl-YPO4, YPO4}.

\begin{table}[]
 \centering
 \caption{\label{barrier} The calculated and experimental values of the energy barrier $\delta$ between JT* and pJT, as well as the zero-phonon line energy difference $\Delta E_{\mathrm{zpl}}$ (in units of meV).  For the ScPO$_{4}$, the $\Delta E_{\mathrm{zpl}}$ between JT and pJT is used as an estimate.}
 \begin{ruledtabular}
 \begin{tabular}{*{14}c}
 \specialrule{0em}{2pt}{2pt}
 %\cline{2-5} \\
$L$PO$_{4}$ &\multicolumn{3}{c}{$\delta$} &\multicolumn{3}{c}{$\Delta E_{\mathrm{zpl}}$}  \\
\specialrule{0em}{2pt}{2pt}
\quad & PBEsol & PBE027 & Exp. &  PBEsol & PBE027 & Exp.  \\
\specialrule{0em}{2pt}{2pt}
\hline 
\specialrule{0em}{2pt}{2pt}
\specialrule{0em}{2pt}{2pt}
Sc &  / & / & / &  /   & 166 & 74~\cite{1984cpl-YPO4}, \\
\quad & \quad & \quad & \quad &  \quad & \quad &  72~\cite{YPO4} \\
 \specialrule{0em}{2pt}{2pt}
Y & 29.3 & 39.7 & 10.0~\cite{1984cpl-YPO4}, &  18 & 148 & 129~\cite{1984cpl-YPO4}, \\
\quad & \quad & \quad & 40.9~\cite{YPO4} &  \quad & \quad &  107~\cite{YPO4} \\
\specialrule{0em}{2pt}{2pt}
Lu & 16.8 & 36.0 & 5.0~\cite{1984cpl-YPO4} & 27 & 108 & 99~\cite{1984cpl-YPO4},   \\
\quad & \quad & \quad & \quad & \quad & \quad &  102~\cite{YPO4}   \\
  \specialrule{0em}{2pt}{2pt}
 \end{tabular}
 \end{ruledtabular}
\end{table}

\section{CONCLUSIONS}
Our calculations reveal that the dual-band emissions of Sb$^{3+}$ in the $L\mathrm{PO_{4}}$ series are primarily governed by two sets of distinct excited-state local minima. These arise from distortions primarily governed by two different set of effective vibration modes  via the Jahn-Teller effect and the pseudo Jahn-Teller effect. The two distinct sets of equilibrium structures for excited states give rise to the experimentally observed ``UV band'' (JT) and the ``visible band'' (pJT). The relative position of the charge transition levels to the host bands significantly impacts the optical properties of the dopant ions. In particular, the differences in the host cations can influence the energy barriers between different excited-state equilibrium structures, thereby affecting the relative intensities of the ``UV band'' and the ``visible band'' at low temperatures. In contrast to 6s$^{2}$ dopant, such as Bi$^{3+}$, the influence of SOC on the potential energy surface of the 5s$^{2}$ dopant, Sb$^{3+}$, remains largely suppressed by the Jahn-Teller effect. This means that SOC is not significant in determining the adiabatic potential energy surface of Sb$^{3+}$, although it is deemed important in lifting the spin-forbidenness in optical transitions. This study demonstrates that the interplay of ligand fields, vibration modes, and spin-orbit interactions may lead to rich phenomena in excited state dynamics.

\begin{acknowledgments}
This work was supported by the National Natural Science Foundation of China (Grants No.\ 12474242 and No.\ 62375255), the Innovation Program for Quantum Science and Technology (Grant No. 2021ZD0302200), and University Science Research Project of Anhui Province (Grant No. KJ2020A0820). 
The numerical calculations were partially performed on the supercomputing system at the Supercomputing Center of the University of Science and Technology of China.

\end{acknowledgments}

\appendix
\section{Clebsch-Gordan Coefficients}

\begin{table}[htbp]
 \caption{Clebsch-Gordan coefficients of E $\otimes$ E in D$_{2d}$ point group }
 \label{CG1}
 \begin{tabular}{cc|cccc}
 \\
 \multicolumn{2}{c|}{E $\otimes$ E} & A$_{1}$ & A$_{2}$ & B$_{1}$ & B$_{2}$ \\
 \hline
 $\gamma_1$ & $\gamma_2$ \\
 \multirow{2}{*}{$u$} & $u$ & $\dfrac{1}{\sqrt{2}}$ & 0 & 0 &$\dfrac{1}{\sqrt{2}}$ \\
 & $v$ & 0 & $\dfrac{1}{\sqrt{2}}$ & $\dfrac{1}{\sqrt{2}}$ & 0 \\
 \quad & \quad & \quad & \quad & \quad & \quad \\
 \multirow{2}{*}{$v$} & $u$ & 0 & $-\dfrac{1}{\sqrt{2}}$ & $\dfrac{1}{\sqrt{2}}$ & 0 \\
 & $v$ & $\dfrac{1}{\sqrt{2}}$ & 0 & 0 &$-\dfrac{1}{\sqrt{2}}$\\ 
 \end{tabular}
\end{table}

\begin{table}[htbp]
 \caption{Clebsch-Gordan coefficients of B$_{1}$ $\otimes$ E in D$_{2d}$ point group}
 \label{CG2}
 \begin{tabular}{cc|cccc}
 \\
 \multicolumn{2}{c|}{B$_{1}$ $\otimes$ E} & \multicolumn{2}{c}{E}\\
 \quad & \quad & $u$ & $v$ \\
 \hline
 $\gamma_1$ & $\gamma_2$ \\
 \multirow{2}{*}{$b_1$} & $u$ & 0 &1 \\
 & $v$ & 1 & 0 \\ 
 \end{tabular}
\end{table}

\begin{table}[htbp]
 \caption{Clebsch-Gordan coefficients of B$_{2}$ $\otimes$ E in D$_{2d}$ point group}
 \label{CG3}
 \begin{tabular}{cc|cccc}
 \\
 \multicolumn{2}{c|}{B$_{2}$ $\otimes$ E} & \multicolumn{2}{c}{E}\\
 \quad & \quad & $u$ & $v$ \\
 \hline
 $\gamma_1$ & $\gamma_2$ \\
 \multirow{2}{*}{$b_2$} & $u$ &1 & 0 \\
 & $v$ & 0 & -1 \\ 
 \end{tabular}
\end{table}
%\nocite{*}

\begin{figure}[htbp]
 \centering
 \begin{overpic}[width=0.50\columnwidth]{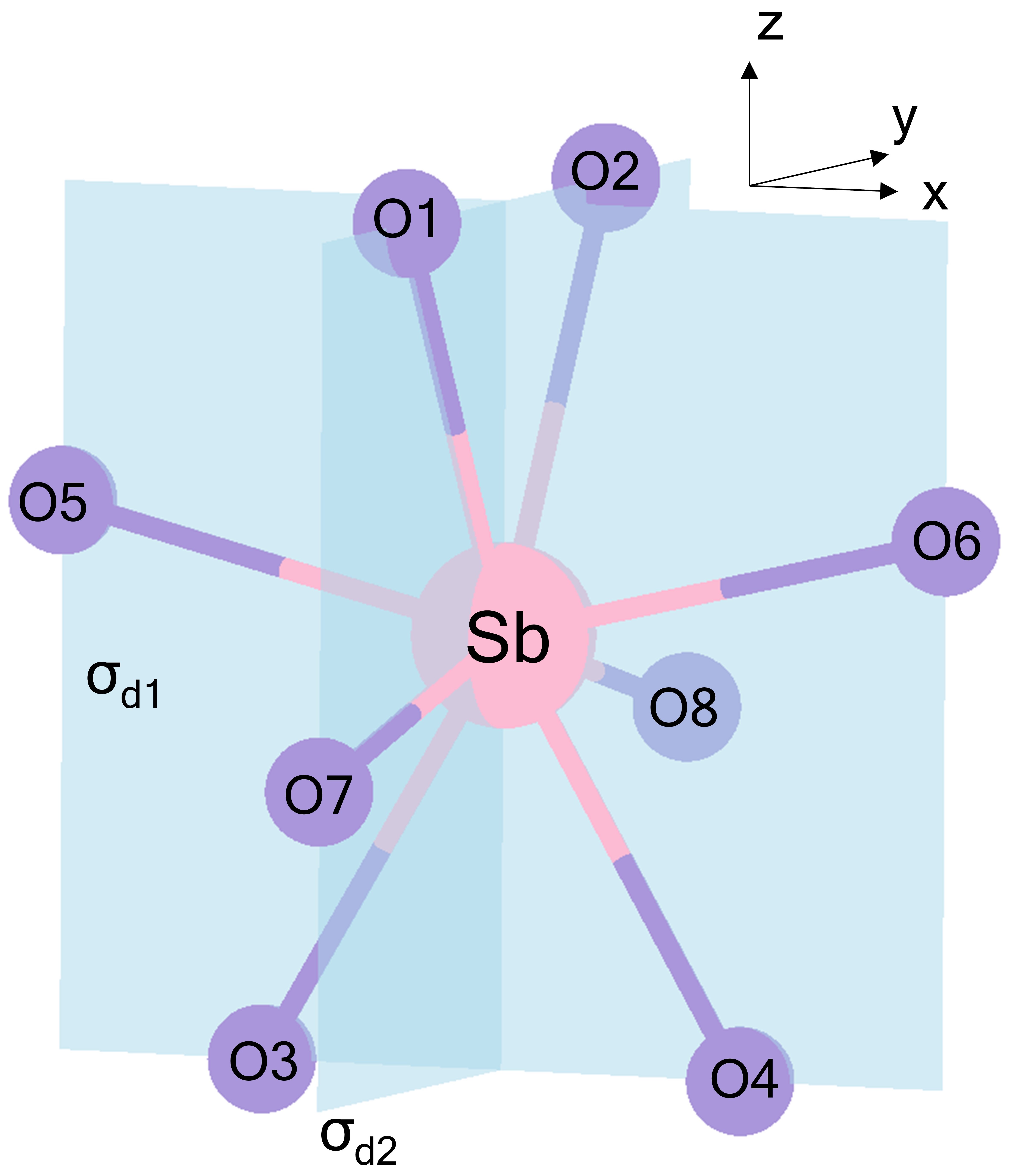}
 \end{overpic} 
 \caption{The SbO$_{8}$ local structure in Sb: $L$PO$_4$ and $\sigma_{d1,2}$ represent two diagonal reflection planes of the point group D$_{2d}$. For convenience, we establish a new coordinate system so that $x$-axis and $y$-axis are normal to the reflection planes $\sigma_{d2}$ and $\sigma_{d1}$, respectively.}
 \label{SbO8}
\end{figure}

\section{\label{vib_mode} Vibrational Modes}

In this section, we list the vibrational modes of [SbO$_{8}$] in Sb: $L$PO$_{4}$. The local structure of [SbO$_{8}$] is shown as in Fig.~\ref{SbO8}

Sb: $
\begin{cases}
& B_2: z_{0} \\

& E: \begin{cases} 
 & x_{0} \\
 & y_{0} \\
\end{cases} \\

\end{cases}
$

\quad \\

$A_{1} : 
\begin{cases} 
 & O_{1z} + O_{2z} - O_{3z} - O_{4z} \\
 & O_{5z} + O_{6z} - O_{7z} - O_{8z} \\
 & O_{1y} - O_{2y} + O_{3x} - O_{4x} \\
 & O_{5x} - O_{6x} + O_{7y} - O_{8y} \\
\end{cases}
$

\quad \\

$A_{2} : 
\begin{cases} 
 & - O_{1x} + O_{2x} + O_{3y} - O_{4y} \\
 & O_{5y} - O_{6y} - O_{7x} + O_{8x} \\
\end{cases}
$

\quad \\

$B_{1} : 
\begin{cases} 
 & O_{1x} - O_{2x} + O_{3y} - O_{4y} \\
 & O_{5y} - O_{6y} + O_{7x} - O_{8x} \\
\end{cases}
$

\quad \\

$B_{2} : 
\begin{cases} 
 & O_{1z} + O_{2z} + O_{3z} + O_{4z} \\
 & O_{5z} + O_{6z} + O_{7z} + O_{8z} \\
 & - O_{1y} + O_{2y} + O_{3x} - O_{4x} \\
 & O_{5x} - O_{6x} - O_{7y} + O_{8y} \\
\end{cases}
$

\quad \\

$E : \begin{cases} 
 & E_1 
 \begin{cases}
 & - O_{1z} + O_{2z} \\
 & O_{3z} - O_{4z} \\
 \end{cases} \\
 \quad \\
 & E_2 
 \begin{cases}
 & O_{5z} - O_{6z} \\
 & - O_{7z} + O_{8z} \\
 \end{cases} \\
 \quad \\
 & E_3 
 \begin{cases}
 & O_{1x} + O_{2x} - O_{3x} - O_{4x} \\
 & - O_{1y} - O_{2y} + O_{3y} + O_{4y} \\
 \end{cases} \\
 \quad \\
 & E_4 
 \begin{cases}
 & O_{5x} + O_{6x} - O_{7x} - O_{8x} \\
 & - O_{5y} - O_{6y} + O_{7y} + O_{8y} \\
 \end{cases} \\
 \quad \\
 & E_5 
 \begin{cases}
 & O_{1x} + O_{2x} + O_{3x} + O_{4x} \\
 & O_{1y} + O_{2y} + O_{3y} + O_{4y} \\
 \end{cases} \\
 \quad \\
 & E_6 
 \begin{cases}
 & O_{5x} + O_{6x} + O_{7x} + O_{8x} \\
 & O_{5y} + O_{6y} + O_{7y} + O_{8y} \\
 \end{cases}
\end{cases}
$

%\subsection{\label{app:subsec}A subsection in an appendix}
\clearpage
%\bibliography{manuscript.bib}% Produces the bibliography via BibTeX.

\begin{thebibliography}{64}%
  \makeatletter
  \providecommand \@ifxundefined [1]{%
   \@ifx{#1\undefined}
  }%
  \providecommand \@ifnum [1]{%
   \ifnum #1\expandafter \@firstoftwo
   \else \expandafter \@secondoftwo
   \fi
  }%
  \providecommand \@ifx [1]{%
   \ifx #1\expandafter \@firstoftwo
   \else \expandafter \@secondoftwo
   \fi
  }%
  \providecommand \natexlab [1]{#1}%
  \providecommand \enquote  [1]{``#1''}%
  \providecommand \bibnamefont  [1]{#1}%
  \providecommand \bibfnamefont [1]{#1}%
  \providecommand \citenamefont [1]{#1}%
  \providecommand \href@noop [0]{\@secondoftwo}%
  \providecommand \href [0]{\begingroup \@sanitize@url \@href}%
  \providecommand \@href[1]{\@@startlink{#1}\@@href}%
  \providecommand \@@href[1]{\endgroup#1\@@endlink}%
  \providecommand \@sanitize@url [0]{\catcode `\\12\catcode `\$12\catcode
    `\&12\catcode `\#12\catcode `\^12\catcode `\_12\catcode `\%12\relax}%
  \providecommand \@@startlink[1]{}%
  \providecommand \@@endlink[0]{}%
  \providecommand \url  [0]{\begingroup\@sanitize@url \@url }%
  \providecommand \@url [1]{\endgroup\@href {#1}{\urlprefix }}%
  \providecommand \urlprefix  [0]{URL }%
  \providecommand \Eprint [0]{\href }%
  \providecommand \doibase [0]{https://doi.org/}%
  \providecommand \selectlanguage [0]{\@gobble}%
  \providecommand \bibinfo  [0]{\@secondoftwo}%
  \providecommand \bibfield  [0]{\@secondoftwo}%
  \providecommand \translation [1]{[#1]}%
  \providecommand \BibitemOpen [0]{}%
  \providecommand \bibitemStop [0]{}%
  \providecommand \bibitemNoStop [0]{.\EOS\space}%
  \providecommand \EOS [0]{\spacefactor3000\relax}%
  \providecommand \BibitemShut  [1]{\csname bibitem#1\endcsname}%
  \let\auto@bib@innerbib\@empty
  %</preamble>
  \bibitem [{\citenamefont {Li}\ \emph {et~al.}(2023)\citenamefont {Li},
    \citenamefont {Liang}, \citenamefont {Xiao}, \citenamefont {Jia},
    \citenamefont {Guo}, \citenamefont {Chen}, \citenamefont {Guo}, \citenamefont
    {Luo}, \citenamefont {Wang}, \citenamefont {Li}, \citenamefont {Rossier},
    \citenamefont {Hauser}, \citenamefont {Linardi}, \citenamefont {Alvianto},
    \citenamefont {Liu}, \citenamefont {Feng},\ and\ \citenamefont
    {Hou}}]{2023NatEnergy-ATO}%
    \BibitemOpen
    \bibfield  {author} {\bibinfo {author} {\bibfnamefont {J.}~\bibnamefont
    {Li}}, \bibinfo {author} {\bibfnamefont {H.}~\bibnamefont {Liang}}, \bibinfo
    {author} {\bibfnamefont {C.}~\bibnamefont {Xiao}}, \bibinfo {author}
    {\bibfnamefont {X.}~\bibnamefont {Jia}}, \bibinfo {author} {\bibfnamefont
    {R.}~\bibnamefont {Guo}}, \bibinfo {author} {\bibfnamefont {J.}~\bibnamefont
    {Chen}}, \bibinfo {author} {\bibfnamefont {X.}~\bibnamefont {Guo}}, \bibinfo
    {author} {\bibfnamefont {R.}~\bibnamefont {Luo}}, \bibinfo {author}
    {\bibfnamefont {X.}~\bibnamefont {Wang}}, \bibinfo {author} {\bibfnamefont
    {M.}~\bibnamefont {Li}}, \bibinfo {author} {\bibfnamefont {M.}~\bibnamefont
    {Rossier}}, \bibinfo {author} {\bibfnamefont {A.}~\bibnamefont {Hauser}},
    \bibinfo {author} {\bibfnamefont {F.}~\bibnamefont {Linardi}}, \bibinfo
    {author} {\bibfnamefont {E.}~\bibnamefont {Alvianto}}, \bibinfo {author}
    {\bibfnamefont {S.}~\bibnamefont {Liu}}, \bibinfo {author} {\bibfnamefont
    {J.}~\bibnamefont {Feng}},\ and\ \bibinfo {author} {\bibfnamefont
    {Y.}~\bibnamefont {Hou}},\ }\bibfield  {title} {\bibinfo {title} {{Enhancing
    the Efficiency and Longevity of Inverted Perovskite Solar Cells with
    Antimony-Doped Tin Oxides}},\ }\href
    {https://doi.org/10.1038/s41560-023-01442-1} {\bibfield  {journal} {\bibinfo
    {journal} {Nature Energy}\ }\textbf {\bibinfo {volume} {9}},\ \bibinfo
    {pages} {308} (\bibinfo {year} {2023})}\BibitemShut {NoStop}%
  \bibitem [{\citenamefont {Wang}\ \emph {et~al.}(2024)\citenamefont {Wang},
    \citenamefont {Loi}, \citenamefont {Cao}, \citenamefont {Feng}, \citenamefont
    {Wei}, \citenamefont {Chen}, \citenamefont {Li}, \citenamefont {Zhu},
    \citenamefont {Lee},\ and\ \citenamefont {Yan}}]{2024AM-Sn}%
    \BibitemOpen
    \bibfield  {author} {\bibinfo {author} {\bibfnamefont {T.}~\bibnamefont
    {Wang}}, \bibinfo {author} {\bibfnamefont {H.-L.}\ \bibnamefont {Loi}},
    \bibinfo {author} {\bibfnamefont {Q.}~\bibnamefont {Cao}}, \bibinfo {author}
    {\bibfnamefont {Z.}~\bibnamefont {Feng}, \bibfnamefont {Guitao~andGuan}},
    \bibinfo {author} {\bibfnamefont {Q.}~\bibnamefont {Wei}}, \bibinfo {author}
    {\bibfnamefont {C.}~\bibnamefont {Chen}}, \bibinfo {author} {\bibfnamefont
    {M.}~\bibnamefont {Li}}, \bibinfo {author} {\bibfnamefont {Y.}~\bibnamefont
    {Zhu}}, \bibinfo {author} {\bibfnamefont {C.-S.}\ \bibnamefont {Lee}},\ and\
    \bibinfo {author} {\bibfnamefont {F.}~\bibnamefont {Yan}},\ }\bibfield
    {title} {\bibinfo {title} {{Counter-Doping Effect by Trivalent Cations in
    Tin-Based Perovskite Solar Cells}},\ }\href
    {https://doi.org/10.1002/adma.202402947} {\bibfield  {journal} {\bibinfo
    {journal} {Advanced Materials}\ ,\ \bibinfo {pages} {2402947}} (\bibinfo
    {year} {2024})}\BibitemShut {NoStop}%
  \bibitem [{\citenamefont {Zhang}\ \emph {et~al.}(2024)\citenamefont {Zhang},
    \citenamefont {Ru}, \citenamefont {Zhou}, \citenamefont {Jia}, \citenamefont
    {Song}, \citenamefont {Liu}, \citenamefont {Zhang}, \citenamefont {Liu},
    \citenamefont {Zhong}, \citenamefont {Yong}, \citenamefont {Panneerselvam},
    \citenamefont {Manna},\ and\ \citenamefont {Lu}}]{2024jacs-antiTQ}%
    \BibitemOpen
    \bibfield  {author} {\bibinfo {author} {\bibfnamefont {B.}~\bibnamefont
    {Zhang}}, \bibinfo {author} {\bibfnamefont {Y.}~\bibnamefont {Ru}}, \bibinfo
    {author} {\bibfnamefont {J.}~\bibnamefont {Zhou}}, \bibinfo {author}
    {\bibfnamefont {J.}~\bibnamefont {Jia}}, \bibinfo {author} {\bibfnamefont
    {H.}~\bibnamefont {Song}}, \bibinfo {author} {\bibfnamefont {Z.}~\bibnamefont
    {Liu}}, \bibinfo {author} {\bibfnamefont {L.}~\bibnamefont {Zhang}}, \bibinfo
    {author} {\bibfnamefont {X.}~\bibnamefont {Liu}}, \bibinfo {author}
    {\bibfnamefont {G.-M.}\ \bibnamefont {Zhong}}, \bibinfo {author}
    {\bibfnamefont {X.}~\bibnamefont {Yong}}, \bibinfo {author} {\bibfnamefont
    {I.~R.}\ \bibnamefont {Panneerselvam}}, \bibinfo {author} {\bibfnamefont
    {L.}~\bibnamefont {Manna}},\ and\ \bibinfo {author} {\bibfnamefont
    {S.}~\bibnamefont {Lu}},\ }\bibfield  {title} {\bibinfo {title} {{A Robust
    Anti-Thermal-Quenching Phosphor Based on Zero-Dimensional Metal Halide
    Rb$_3$InCl$_6$:xSb$^{3+}$}},\ }\href {https://doi.org/10.1021/jacs.3c14137}
    {\bibfield  {journal} {\bibinfo  {journal} {Journal of the American Chemical
    Society}\ }\textbf {\bibinfo {volume} {146}},\ \bibinfo {pages} {7658}
    (\bibinfo {year} {2024})}\BibitemShut {NoStop}%
  \bibitem [{\citenamefont {Zeng}\ \emph {et~al.}(2020)\citenamefont {Zeng},
    \citenamefont {Zhang}, \citenamefont {Xue}, \citenamefont {Ke}, \citenamefont
    {Zhao}, \citenamefont {Huang}, \citenamefont {Wei}, \citenamefont {Zhou},\
    and\ \citenamefont {Zou}}]{2020jpcl-Sb-NaIn}%
    \BibitemOpen
    \bibfield  {author} {\bibinfo {author} {\bibfnamefont {R.}~\bibnamefont
    {Zeng}}, \bibinfo {author} {\bibfnamefont {L.}~\bibnamefont {Zhang}},
    \bibinfo {author} {\bibfnamefont {Y.}~\bibnamefont {Xue}}, \bibinfo {author}
    {\bibfnamefont {B.}~\bibnamefont {Ke}}, \bibinfo {author} {\bibfnamefont
    {Z.}~\bibnamefont {Zhao}}, \bibinfo {author} {\bibfnamefont {D.}~\bibnamefont
    {Huang}}, \bibinfo {author} {\bibfnamefont {Q.}~\bibnamefont {Wei}}, \bibinfo
    {author} {\bibfnamefont {W.}~\bibnamefont {Zhou}},\ and\ \bibinfo {author}
    {\bibfnamefont {B.}~\bibnamefont {Zou}},\ }\bibfield  {title} {\bibinfo
    {title} {{Highly Efficient Blue Emission from Self-Trapped Excitons in Stable
    Sb$^{3+}$-Doped Cs$_2$NaInCl$_6$ Double Perovskites}},\ }\href
    {https://doi.org/10.1021/acs.jpclett.0c00330} {\bibfield  {journal} {\bibinfo
     {journal} {The Journal of Physical Chemistry Letters}\ }\textbf {\bibinfo
    {volume} {11}},\ \bibinfo {pages} {2053} (\bibinfo {year}
    {2020})}\BibitemShut {NoStop}%
  \bibitem [{\citenamefont {Li}\ \emph {et~al.}(2019)\citenamefont {Li},
    \citenamefont {Li}, \citenamefont {Liang}, \citenamefont {Zhou},
    \citenamefont {Wang},\ and\ \citenamefont {Xie}}]{2019CM-organic-Sb}%
    \BibitemOpen
    \bibfield  {author} {\bibinfo {author} {\bibfnamefont {Z.}~\bibnamefont
    {Li}}, \bibinfo {author} {\bibfnamefont {Y.}~\bibnamefont {Li}}, \bibinfo
    {author} {\bibfnamefont {P.}~\bibnamefont {Liang}}, \bibinfo {author}
    {\bibfnamefont {T.}~\bibnamefont {Zhou}}, \bibinfo {author} {\bibfnamefont
    {L.}~\bibnamefont {Wang}},\ and\ \bibinfo {author} {\bibfnamefont {R.-J.}\
    \bibnamefont {Xie}},\ }\bibfield  {title} {\bibinfo {title} {{Dual-Band
    Luminescent Lead-Free Antimony Chloride Halides with Near-Unity
    Photoluminescence Quantum Efficiency}},\ }\href
    {https://doi.org/10.1021/acs.chemmater.9b02935} {\bibfield  {journal}
    {\bibinfo  {journal} {Chemistry of Materials}\ }\textbf {\bibinfo {volume}
    {31}},\ \bibinfo {pages} {9363} (\bibinfo {year} {2019})}\BibitemShut
    {NoStop}%
  \bibitem [{\citenamefont {Jing}\ \emph {et~al.}(2020)\citenamefont {Jing},
    \citenamefont {Liu}, \citenamefont {Jiang}, \citenamefont {Molokeev},
    \citenamefont {Lin},\ and\ \citenamefont {Xia}}]{2020CM-CsInCl5}%
    \BibitemOpen
    \bibfield  {author} {\bibinfo {author} {\bibfnamefont {Y.}~\bibnamefont
    {Jing}}, \bibinfo {author} {\bibfnamefont {Y.}~\bibnamefont {Liu}}, \bibinfo
    {author} {\bibfnamefont {X.}~\bibnamefont {Jiang}}, \bibinfo {author}
    {\bibfnamefont {M.~S.}\ \bibnamefont {Molokeev}}, \bibinfo {author}
    {\bibfnamefont {Z.}~\bibnamefont {Lin}},\ and\ \bibinfo {author}
    {\bibfnamefont {Z.}~\bibnamefont {Xia}},\ }\bibfield  {title} {\bibinfo
    {title} {{Sb$^{3+}$ Dopant and Halogen Substitution Triggered Highly
    Efficient and Tunable Emission in Lead-Free Metal Halide Single Crystals}},\
    }\href {https://doi.org/10.1021/acs.chemmater.0c01708} {\bibfield  {journal}
    {\bibinfo  {journal} {Chemistry of Materials}\ }\textbf {\bibinfo {volume}
    {32}},\ \bibinfo {pages} {5327} (\bibinfo {year} {2020})}\BibitemShut
    {NoStop}%
  \bibitem [{\citenamefont {Su}\ \emph {et~al.}(2022)\citenamefont {Su},
    \citenamefont {Geng}, \citenamefont {Xiao},\ and\ \citenamefont
    {Xia}}]{Sb-pair-organic}%
    \BibitemOpen
    \bibfield  {author} {\bibinfo {author} {\bibfnamefont {B.}~\bibnamefont
    {Su}}, \bibinfo {author} {\bibfnamefont {S.}~\bibnamefont {Geng}}, \bibinfo
    {author} {\bibfnamefont {Z.}~\bibnamefont {Xiao}},\ and\ \bibinfo {author}
    {\bibfnamefont {Z.}~\bibnamefont {Xia}},\ }\bibfield  {title} {\bibinfo
    {title} {{Highly Distorted Antimony(III) Chloride [Sb$_2$Cl$_8$]$^{2-}$
    Dimers for Near-Infrared Luminescence up to 1070nm}},\ }\href
    {https://doi.org/10.1002/anie.202208881} {\bibfield  {journal} {\bibinfo
    {journal} {Angewandte Chemie International Edition}\ }\textbf {\bibinfo
    {volume} {61}},\ \bibinfo {pages} {e202208881} (\bibinfo {year}
    {2022})}\BibitemShut {NoStop}%
  \bibitem [{\citenamefont {Su}\ \emph {et~al.}(2021)\citenamefont {Su},
    \citenamefont {Li}, \citenamefont {Song},\ and\ \citenamefont
    {Xia}}]{2021AFM-Zn}%
    \BibitemOpen
    \bibfield  {author} {\bibinfo {author} {\bibfnamefont {B.}~\bibnamefont
    {Su}}, \bibinfo {author} {\bibfnamefont {M.}~\bibnamefont {Li}}, \bibinfo
    {author} {\bibfnamefont {E.}~\bibnamefont {Song}},\ and\ \bibinfo {author}
    {\bibfnamefont {Z.}~\bibnamefont {Xia}},\ }\bibfield  {title} {\bibinfo
    {title} {{Sb$^{3+}$-Doping in Cesium Zinc Halides Single Crystals Enabling
    High-Efficiency Near-Infrared Emission}},\ }\href
    {https://doi.org/https://doi.org/10.1002/adfm.202105316} {\bibfield
    {journal} {\bibinfo  {journal} {Advanced Functional Materials}\ }\textbf
    {\bibinfo {volume} {31}},\ \bibinfo {pages} {2105316} (\bibinfo {year}
    {2021})}\BibitemShut {NoStop}%
  \bibitem [{\citenamefont {Jin}\ \emph {et~al.}(2022)\citenamefont {Jin},
    \citenamefont {Peng}, \citenamefont {Xu}, \citenamefont {Han}, \citenamefont
    {Zhang}, \citenamefont {Yang},\ and\ \citenamefont {Xia}}]{2022CM-xia}%
    \BibitemOpen
    \bibfield  {author} {\bibinfo {author} {\bibfnamefont {J.-C.}\ \bibnamefont
    {Jin}}, \bibinfo {author} {\bibfnamefont {Y.}~\bibnamefont {Peng}}, \bibinfo
    {author} {\bibfnamefont {Y.}~\bibnamefont {Xu}}, \bibinfo {author}
    {\bibfnamefont {K.}~\bibnamefont {Han}}, \bibinfo {author} {\bibfnamefont
    {A.-R.}\ \bibnamefont {Zhang}}, \bibinfo {author} {\bibfnamefont {X.-B.}\
    \bibnamefont {Yang}},\ and\ \bibinfo {author} {\bibfnamefont
    {Z.}~\bibnamefont {Xia}},\ }\bibfield  {title} {\bibinfo {title} {{Bright
    Green Emission from Self-Trapped Excitons Triggered by Sb$^{3+}$ Doping in
    Rb$_4$CdCl$_6$}},\ }\href {https://doi.org/10.1021/acs.chemmater.2c01254}
    {\bibfield  {journal} {\bibinfo  {journal} {Chemistry of Materials}\ }\textbf
    {\bibinfo {volume} {34}},\ \bibinfo {pages} {5717} (\bibinfo {year}
    {2022})}\BibitemShut {NoStop}%
  \bibitem [{\citenamefont {Wolfert}\ and\ \citenamefont
    {Blasse}(1985)}]{1985-CsNaLaCl}%
    \BibitemOpen
    \bibfield  {author} {\bibinfo {author} {\bibfnamefont {A.}~\bibnamefont
    {Wolfert}}\ and\ \bibinfo {author} {\bibfnamefont {G.}~\bibnamefont
    {Blasse}},\ }\bibfield  {title} {\bibinfo {title} {{Luminescence of
    Bi$^{3+}$- Doped Crystals of Cs$_2$NaYBr$_6$ and Cs$_2$NaLaCl$_6$}},\ }\href
    {https://doi.org/10.1016/0022-4596(85)90310-x} {\bibfield  {journal}
    {\bibinfo  {journal} {Journal of Solid State Chemistry}\ }\textbf {\bibinfo
    {volume} {59}},\ \bibinfo {pages} {133} (\bibinfo {year} {1985})}\BibitemShut
    {NoStop}%
  \bibitem [{\citenamefont {Majher}\ \emph {et~al.}(2020)\citenamefont {Majher},
    \citenamefont {Gray}, \citenamefont {Liu}, \citenamefont {Holzapfel},\ and\
    \citenamefont {Woodward}}]{2020IC-RbIn}%
    \BibitemOpen
    \bibfield  {author} {\bibinfo {author} {\bibfnamefont {J.~D.}\ \bibnamefont
    {Majher}}, \bibinfo {author} {\bibfnamefont {M.~B.}\ \bibnamefont {Gray}},
    \bibinfo {author} {\bibfnamefont {T.}~\bibnamefont {Liu}}, \bibinfo {author}
    {\bibfnamefont {N.~P.}\ \bibnamefont {Holzapfel}},\ and\ \bibinfo {author}
    {\bibfnamefont {P.~M.}\ \bibnamefont {Woodward}},\ }\bibfield  {title}
    {\bibinfo {title} {{Rb$_3$InCl$_6$: A Monoclinic Double Perovskite Derivative
    with Bright Sb$^{3+}$-Activated Photoluminescence}},\ }\href
    {https://doi.org/10.1021/acs.inorgchem.0c02248} {\bibfield  {journal}
    {\bibinfo  {journal} {Inorganic Chemistry}\ }\textbf {\bibinfo {volume}
    {59}},\ \bibinfo {pages} {14478} (\bibinfo {year} {2020})}\BibitemShut
    {NoStop}%
  \bibitem [{\citenamefont {Noculak}\ \emph {et~al.}(2020)\citenamefont
    {Noculak}, \citenamefont {Morad}, \citenamefont {McCall}, \citenamefont
    {Yakunin}, \citenamefont {Shynkarenko}, \citenamefont {Wörle},\ and\
    \citenamefont {Kovalenko}}]{2020CM-Sb-KIn}%
    \BibitemOpen
    \bibfield  {author} {\bibinfo {author} {\bibfnamefont {A.}~\bibnamefont
    {Noculak}}, \bibinfo {author} {\bibfnamefont {V.}~\bibnamefont {Morad}},
    \bibinfo {author} {\bibfnamefont {K.~M.}\ \bibnamefont {McCall}}, \bibinfo
    {author} {\bibfnamefont {S.}~\bibnamefont {Yakunin}}, \bibinfo {author}
    {\bibfnamefont {Y.}~\bibnamefont {Shynkarenko}}, \bibinfo {author}
    {\bibfnamefont {M.}~\bibnamefont {Wörle}},\ and\ \bibinfo {author}
    {\bibfnamefont {M.~V.}\ \bibnamefont {Kovalenko}},\ }\bibfield  {title}
    {\bibinfo {title} {{Bright Blue and Green Luminescence of Sb(III) in Double
    Perovskite Cs$_2$MInCl$_6$ (M = Na, K) Matrices}},\ }\href
    {https://doi.org/10.1021/acs.chemmater.0c01004} {\bibfield  {journal}
    {\bibinfo  {journal} {Chemistry of Materials}\ }\textbf {\bibinfo {volume}
    {32}},\ \bibinfo {pages} {5118} (\bibinfo {year} {2020})}\BibitemShut
    {NoStop}%
  \bibitem [{\citenamefont {Locardi}\ \emph {et~al.}(2021)\citenamefont
    {Locardi}, \citenamefont {Samoli}, \citenamefont {Martinelli}, \citenamefont
    {Erdem}, \citenamefont {Magalhaes}, \citenamefont {Bals},\ and\ \citenamefont
    {Hens}}]{2021Nano-Cs2CdCl4}%
    \BibitemOpen
    \bibfield  {author} {\bibinfo {author} {\bibfnamefont {F.}~\bibnamefont
    {Locardi}}, \bibinfo {author} {\bibfnamefont {M.}~\bibnamefont {Samoli}},
    \bibinfo {author} {\bibfnamefont {A.}~\bibnamefont {Martinelli}}, \bibinfo
    {author} {\bibfnamefont {O.}~\bibnamefont {Erdem}}, \bibinfo {author}
    {\bibfnamefont {D.~V.}\ \bibnamefont {Magalhaes}}, \bibinfo {author}
    {\bibfnamefont {S.}~\bibnamefont {Bals}},\ and\ \bibinfo {author}
    {\bibfnamefont {Z.}~\bibnamefont {Hens}},\ }\bibfield  {title} {\bibinfo
    {title} {{Cyan Emission in Two-Dimensional Colloidal Cs$_2$CdCl$_4$:Sb$^{3+}$
    Ruddlesden-Popper Phase Nanoplatelets}},\ }\href
    {https://doi.org/10.1021/acsnano.1c05684} {\bibfield  {journal} {\bibinfo
    {journal} {ACS Nano}\ }\textbf {\bibinfo {volume} {15}},\ \bibinfo {pages}
    {17729} (\bibinfo {year} {2021})}\BibitemShut {NoStop}%
  \bibitem [{\citenamefont {Oomen}\ \emph {et~al.}(1986)\citenamefont {Oomen},
    \citenamefont {Smit},\ and\ \citenamefont {Blasse}}]{1986-Sb-Cs2NaMCl6}%
    \BibitemOpen
    \bibfield  {author} {\bibinfo {author} {\bibfnamefont {E.~W. J.~L.}\
    \bibnamefont {Oomen}}, \bibinfo {author} {\bibfnamefont {W.~M.~A.}\
    \bibnamefont {Smit}},\ and\ \bibinfo {author} {\bibfnamefont
    {G.}~\bibnamefont {Blasse}},\ }\bibfield  {title} {\bibinfo {title} {{On the
    Luminescence of Sb$^{3+}$ in Cs$_2$NaMCl$_6$ (with M=Sc,Y,La): A Model System
    for the Study of Trivalent s$^{2}$ Ions}},\ }\href
    {https://doi.org/10.1088/0022-3719/19/17/020} {\bibfield  {journal} {\bibinfo
     {journal} {Journal of Physics C: Solid State Physics}\ }\textbf {\bibinfo
    {volume} {19}},\ \bibinfo {pages} {3263} (\bibinfo {year}
    {1986})}\BibitemShut {NoStop}%
  \bibitem [{\citenamefont {Jing}\ \emph {et~al.}(2019)\citenamefont {Jing},
    \citenamefont {Liu}, \citenamefont {Zhao},\ and\ \citenamefont
    {Xia}}]{2019jpcl-xia}%
    \BibitemOpen
    \bibfield  {author} {\bibinfo {author} {\bibfnamefont {Y.}~\bibnamefont
    {Jing}}, \bibinfo {author} {\bibfnamefont {Y.}~\bibnamefont {Liu}}, \bibinfo
    {author} {\bibfnamefont {J.}~\bibnamefont {Zhao}},\ and\ \bibinfo {author}
    {\bibfnamefont {Z.}~\bibnamefont {Xia}},\ }\bibfield  {title} {\bibinfo
    {title} {{Sb$^{3+}$ Doping-Induced Triplet Self-Trapped Excitons Emission in
    Lead-Free Cs$_2$SnCl$_6$ Nanocrystals}},\ }\href
    {https://doi.org/10.1021/acs.jpclett.9b03035} {\bibfield  {journal} {\bibinfo
     {journal} {The Journal of Physical Chemistry Letters}\ }\textbf {\bibinfo
    {volume} {10}},\ \bibinfo {pages} {7439} (\bibinfo {year}
    {2019})}\BibitemShut {NoStop}%
  \bibitem [{\citenamefont {Chen}\ \emph {et~al.}(2021)\citenamefont {Chen},
    \citenamefont {Guo}, \citenamefont {Wang}, \citenamefont {Liu}, \citenamefont
    {Wei}, \citenamefont {Wang}, \citenamefont {Rogach}, \citenamefont {Xing},
    \citenamefont {Shi},\ and\ \citenamefont {Wang}}]{2021jacs-Sb-CZC}%
    \BibitemOpen
    \bibfield  {author} {\bibinfo {author} {\bibfnamefont {B.}~\bibnamefont
    {Chen}}, \bibinfo {author} {\bibfnamefont {Y.}~\bibnamefont {Guo}}, \bibinfo
    {author} {\bibfnamefont {Y.}~\bibnamefont {Wang}}, \bibinfo {author}
    {\bibfnamefont {Z.}~\bibnamefont {Liu}}, \bibinfo {author} {\bibfnamefont
    {Q.}~\bibnamefont {Wei}}, \bibinfo {author} {\bibfnamefont {S.}~\bibnamefont
    {Wang}}, \bibinfo {author} {\bibfnamefont {A.~L.}\ \bibnamefont {Rogach}},
    \bibinfo {author} {\bibfnamefont {G.}~\bibnamefont {Xing}}, \bibinfo {author}
    {\bibfnamefont {P.}~\bibnamefont {Shi}},\ and\ \bibinfo {author}
    {\bibfnamefont {F.}~\bibnamefont {Wang}},\ }\bibfield  {title} {\bibinfo
    {title} {{Multiexcitonic Emission in Zero-Dimensional
    Cs$_2$ZrCl$_6$:Sb$^{3+}$ Perovskite Crystals}},\ }\href
    {https://doi.org/10.1021/jacs.1c07537} {\bibfield  {journal} {\bibinfo
    {journal} {Journal of the American Chemical Society}\ }\textbf {\bibinfo
    {volume} {143}},\ \bibinfo {pages} {17599} (\bibinfo {year}
    {2021})}\BibitemShut {NoStop}%
  \bibitem [{\citenamefont {Liu}\ \emph {et~al.}(2021{\natexlab{a}})\citenamefont
    {Liu}, \citenamefont {Zhang}, \citenamefont {Li},\ and\ \citenamefont
    {Liu}}]{Sb-Hf-2021CC}%
    \BibitemOpen
    \bibfield  {author} {\bibinfo {author} {\bibfnamefont {R.}~\bibnamefont
    {Liu}}, \bibinfo {author} {\bibfnamefont {W.}~\bibnamefont {Zhang}}, \bibinfo
    {author} {\bibfnamefont {G.}~\bibnamefont {Li}},\ and\ \bibinfo {author}
    {\bibfnamefont {W.}~\bibnamefont {Liu}},\ }\bibfield  {title} {\bibinfo
    {title} {{Excitation Wavelength Tunable White Light Emission in
    Vacancy-Ordered Double Perovskite}},\ }\href
    {https://doi.org/10.1039/d1cc03208a} {\bibfield  {journal} {\bibinfo
    {journal} {Chemical Communications}\ }\textbf {\bibinfo {volume} {57}},\
    \bibinfo {pages} {10943} (\bibinfo {year} {2021}{\natexlab{a}})}\BibitemShut
    {NoStop}%
  \bibitem [{\citenamefont {Meng}\ \emph {et~al.}(2022)\citenamefont {Meng},
    \citenamefont {Wei}, \citenamefont {Lin}, \citenamefont {Huang},
    \citenamefont {Ge}, \citenamefont {Yu},\ and\ \citenamefont
    {Zou}}]{2022IC-zou}%
    \BibitemOpen
    \bibfield  {author} {\bibinfo {author} {\bibfnamefont {X.}~\bibnamefont
    {Meng}}, \bibinfo {author} {\bibfnamefont {Q.}~\bibnamefont {Wei}}, \bibinfo
    {author} {\bibfnamefont {W.}~\bibnamefont {Lin}}, \bibinfo {author}
    {\bibfnamefont {T.}~\bibnamefont {Huang}}, \bibinfo {author} {\bibfnamefont
    {S.}~\bibnamefont {Ge}}, \bibinfo {author} {\bibfnamefont {Z.}~\bibnamefont
    {Yu}},\ and\ \bibinfo {author} {\bibfnamefont {B.}~\bibnamefont {Zou}},\
    }\bibfield  {title} {\bibinfo {title} {{Efficient Yellow Self-Trapped Exciton
    Emission in Sb$^{3+}$-Doped RbCdCl$_3$ Metal Halides}},\ }\href
    {https://doi.org/10.1021/acs.inorgchem.2c00667} {\bibfield  {journal}
    {\bibinfo  {journal} {Inorganic Chemistry}\ }\textbf {\bibinfo {volume}
    {61}},\ \bibinfo {pages} {7143} (\bibinfo {year} {2022})}\BibitemShut
    {NoStop}%
  \bibitem [{\citenamefont {Ge}\ \emph {et~al.}(2023)\citenamefont {Ge},
    \citenamefont {Peng}, \citenamefont {Wei}, \citenamefont {Shen},
    \citenamefont {Huang}, \citenamefont {Liang}, \citenamefont {Zhao},\ and\
    \citenamefont {Zou}}]{2023AOM-Sb-CsCdCl3}%
    \BibitemOpen
    \bibfield  {author} {\bibinfo {author} {\bibfnamefont {S.}~\bibnamefont
    {Ge}}, \bibinfo {author} {\bibfnamefont {H.}~\bibnamefont {Peng}}, \bibinfo
    {author} {\bibfnamefont {Q.}~\bibnamefont {Wei}}, \bibinfo {author}
    {\bibfnamefont {X.}~\bibnamefont {Shen}}, \bibinfo {author} {\bibfnamefont
    {W.}~\bibnamefont {Huang}}, \bibinfo {author} {\bibfnamefont
    {W.}~\bibnamefont {Liang}}, \bibinfo {author} {\bibfnamefont
    {J.}~\bibnamefont {Zhao}},\ and\ \bibinfo {author} {\bibfnamefont
    {B.}~\bibnamefont {Zou}},\ }\bibfield  {title} {\bibinfo {title} {{Realizing
    Color-Tunable and Time-Dependent Ultralong Afterglow Emission in
    Antimony-Doped CsCdCl$_3$ Metal Halide for Advanced Anti-Counterfeiting and
    Information Encryption}},\ }\href
    {https://doi.org/https://doi.org/10.1002/adom.202300323} {\bibfield
    {journal} {\bibinfo  {journal} {Advanced Optical Materials}\ }\textbf
    {\bibinfo {volume} {11}},\ \bibinfo {pages} {2300323} (\bibinfo {year}
    {2023})}\BibitemShut {NoStop}%
  \bibitem [{\citenamefont {Zhou}\ \emph {et~al.}(2022)\citenamefont {Zhou},
    \citenamefont {Shi}, \citenamefont {Li}, \citenamefont {Sun}, \citenamefont
    {Ji}, \citenamefont {Deng},\ and\ \citenamefont {Wang}}]{2022Ceramics}%
    \BibitemOpen
    \bibfield  {author} {\bibinfo {author} {\bibfnamefont {J.}~\bibnamefont
    {Zhou}}, \bibinfo {author} {\bibfnamefont {C.}~\bibnamefont {Shi}}, \bibinfo
    {author} {\bibfnamefont {X.}~\bibnamefont {Li}}, \bibinfo {author}
    {\bibfnamefont {Z.}~\bibnamefont {Sun}}, \bibinfo {author} {\bibfnamefont
    {Y.}~\bibnamefont {Ji}}, \bibinfo {author} {\bibfnamefont {J.}~\bibnamefont
    {Deng}},\ and\ \bibinfo {author} {\bibfnamefont {B.}~\bibnamefont {Wang}},\
    }\bibfield  {title} {\bibinfo {title} {{A Heterovalent Doping Strategy
    Induced Efficient Cyan Emission in Sb$^{3+}$-Doped CsCdCl$_{3}$ Perovskite
    Microcrystal for Solid State Lighting}},\ }\href
    {https://doi.org/10.1016/j.ceramint.2022.06.139} {\bibfield  {journal}
    {\bibinfo  {journal} {Ceramics International}\ }\textbf {\bibinfo {volume}
    {48}},\ \bibinfo {pages} {28327} (\bibinfo {year} {2022})}\BibitemShut
    {NoStop}%
  \bibitem [{\citenamefont {Jing}\ \emph {et~al.}(2021)\citenamefont {Jing},
    \citenamefont {Liu}, \citenamefont {Li},\ and\ \citenamefont
    {Xia}}]{2021AOM-Sb-Xia}%
    \BibitemOpen
    \bibfield  {author} {\bibinfo {author} {\bibfnamefont {Y.}~\bibnamefont
    {Jing}}, \bibinfo {author} {\bibfnamefont {Y.}~\bibnamefont {Liu}}, \bibinfo
    {author} {\bibfnamefont {M.}~\bibnamefont {Li}},\ and\ \bibinfo {author}
    {\bibfnamefont {Z.}~\bibnamefont {Xia}},\ }\bibfield  {title} {\bibinfo
    {title} {{Photoluminescence of Singlet/Triplet Self-Trapped Excitons in
    Sb$^{3+}$-Based Metal Halides}},\ }\href
    {https://doi.org/10.1002/adom.202002213} {\bibfield  {journal} {\bibinfo
    {journal} {Advanced Optical Materials}\ }\textbf {\bibinfo {volume} {9}},\
    \bibinfo {pages} {2002213} (\bibinfo {year} {2021})}\BibitemShut {NoStop}%
  \bibitem [{\citenamefont {Arfin}\ and\ \citenamefont
    {Nag}(2021)}]{Nag-2021jpcl}%
    \BibitemOpen
    \bibfield  {author} {\bibinfo {author} {\bibfnamefont {H.}~\bibnamefont
    {Arfin}}\ and\ \bibinfo {author} {\bibfnamefont {A.}~\bibnamefont {Nag}},\
    }\bibfield  {title} {\bibinfo {title} {{Origin of Luminescence in Sb$^{3+}$-
    and Bi$^{3+}$-Doped Cs$_2$SnCl$_6$ Perovskites: Excited State Relaxation and
    Spin-Orbit Coupling}},\ }\href {https://doi.org/10.1021/acs.jpclett.1c02973}
    {\bibfield  {journal} {\bibinfo  {journal} {The Journal of Physical Chemistry
    Letters}\ }\textbf {\bibinfo {volume} {12}},\ \bibinfo {pages} {10002}
    (\bibinfo {year} {2021})}\BibitemShut {NoStop}%
  \bibitem [{\citenamefont {Gong}\ \emph {et~al.}(2022)\citenamefont {Gong},
    \citenamefont {Zheng}, \citenamefont {Huang}, \citenamefont {Cheng},
    \citenamefont {Zhang}, \citenamefont {Zhang}, \citenamefont {Han},\ and\
    \citenamefont {Chen}}]{Gong2022}%
    \BibitemOpen
    \bibfield  {author} {\bibinfo {author} {\bibfnamefont {Z.}~\bibnamefont
    {Gong}}, \bibinfo {author} {\bibfnamefont {W.}~\bibnamefont {Zheng}},
    \bibinfo {author} {\bibfnamefont {P.}~\bibnamefont {Huang}}, \bibinfo
    {author} {\bibfnamefont {X.}~\bibnamefont {Cheng}}, \bibinfo {author}
    {\bibfnamefont {W.}~\bibnamefont {Zhang}}, \bibinfo {author} {\bibfnamefont
    {M.}~\bibnamefont {Zhang}}, \bibinfo {author} {\bibfnamefont
    {S.}~\bibnamefont {Han}},\ and\ \bibinfo {author} {\bibfnamefont
    {X.}~\bibnamefont {Chen}},\ }\bibfield  {title} {\bibinfo {title} {{Highly
    Efficient Sb$^{3+}$ Emitters in 0D Cesium Indium Chloride Nanocrystals with
    Switchable Photoluminescence Through Water-Triggered Structural
    Transformation}},\ }\href {https://doi.org/10.1016/j.nantod.2022.101460}
    {\bibfield  {journal} {\bibinfo  {journal} {Nano Today}\ }\textbf {\bibinfo
    {volume} {44}},\ \bibinfo {pages} {101460} (\bibinfo {year}
    {2022})}\BibitemShut {NoStop}%
  \bibitem [{\citenamefont {Zhao}\ \emph {et~al.}(2024)\citenamefont {Zhao},
    \citenamefont {Liu}, \citenamefont {Han}, \citenamefont {Chen}, \citenamefont
    {Jia}, \citenamefont {Zhang}, \citenamefont {Lian}, \citenamefont {Wu},
    \citenamefont {Li},\ and\ \citenamefont {Shi}}]{Zhao2024}%
    \BibitemOpen
    \bibfield  {author} {\bibinfo {author} {\bibfnamefont {D.}~\bibnamefont
    {Zhao}}, \bibinfo {author} {\bibfnamefont {Y.}~\bibnamefont {Liu}}, \bibinfo
    {author} {\bibfnamefont {Y.}~\bibnamefont {Han}}, \bibinfo {author}
    {\bibfnamefont {X.}~\bibnamefont {Chen}}, \bibinfo {author} {\bibfnamefont
    {M.}~\bibnamefont {Jia}}, \bibinfo {author} {\bibfnamefont {J.}~\bibnamefont
    {Zhang}}, \bibinfo {author} {\bibfnamefont {L.}~\bibnamefont {Lian}},
    \bibinfo {author} {\bibfnamefont {D.}~\bibnamefont {Wu}}, \bibinfo {author}
    {\bibfnamefont {X.}~\bibnamefont {Li}},\ and\ \bibinfo {author}
    {\bibfnamefont {Z.}~\bibnamefont {Shi}},\ }\bibfield  {title} {\bibinfo
    {title} {{Tunable White Light Emission in Antimony Doped Vacancy-Ordered
    Double Perovskite Cs$_2$ZrCl$_6$ nanocrystals}},\ }\href
    {https://doi.org/10.1016/j.jlumin.2024.120548} {\bibfield  {journal}
    {\bibinfo  {journal} {Journal of Luminescence}\ }\textbf {\bibinfo {volume}
    {270}},\ \bibinfo {pages} {120548} (\bibinfo {year} {2024})}\BibitemShut
    {NoStop}%
  \bibitem [{\citenamefont {Fukuda}(1970)}]{1970Fukuda-prb}%
    \BibitemOpen
    \bibfield  {author} {\bibinfo {author} {\bibfnamefont {A.}~\bibnamefont
    {Fukuda}},\ }\bibfield  {title} {\bibinfo {title} {{Jahn-Teller Effect on the
    Structure of the Emission Produced by Excitation in the $A$ Band of KI:
    Tl-Type Phosphors. Two Kinds of Minima on the $\Gamma_{4}^{-}$ ($^{3}T_{1u}$)
    Adiabatic Potential-Energy Surface}},\ }\href
    {https://doi.org/10.1103/physrevb.1.4161} {\bibfield  {journal} {\bibinfo
    {journal} {Physical Review B}\ }\textbf {\bibinfo {volume} {1}},\ \bibinfo
    {pages} {4161} (\bibinfo {year} {1970})}\BibitemShut {NoStop}%
  \bibitem [{\citenamefont {Liu}\ \emph {et~al.}(2021{\natexlab{b}})\citenamefont
    {Liu}, \citenamefont {Duan}, \citenamefont {Tanner}, \citenamefont {Ma},\
    and\ \citenamefont {Yin}}]{lmz-2021jpcc}%
    \BibitemOpen
    \bibfield  {author} {\bibinfo {author} {\bibfnamefont {M.}~\bibnamefont
    {Liu}}, \bibinfo {author} {\bibfnamefont {C.-K.}\ \bibnamefont {Duan}},
    \bibinfo {author} {\bibfnamefont {P.~A.}\ \bibnamefont {Tanner}}, \bibinfo
    {author} {\bibfnamefont {C.-G.}\ \bibnamefont {Ma}},\ and\ \bibinfo {author}
    {\bibfnamefont {M.}~\bibnamefont {Yin}},\ }\bibfield  {title} {\bibinfo
    {title} {{Rationalizing the Photoluminescence of Bi$^{3+}$ and Sb$^{3+}$ in
    Double Perovskite Halide Crystals}},\ }\href
    {https://doi.org/10.1021/acs.jpcc.1c09069} {\bibfield  {journal} {\bibinfo
    {journal} {The Journal of Physical Chemistry C}\ }\textbf {\bibinfo {volume}
    {125}},\ \bibinfo {pages} {26670} (\bibinfo {year}
    {2021}{\natexlab{b}})}\BibitemShut {NoStop}%
  \bibitem [{\citenamefont {Liu}\ \emph {et~al.}(2022)\citenamefont {Liu},
    \citenamefont {Duan}, \citenamefont {Tanner}, \citenamefont {Ma},
    \citenamefont {Wei},\ and\ \citenamefont {Yin}}]{2022prb-lmz}%
    \BibitemOpen
    \bibfield  {author} {\bibinfo {author} {\bibfnamefont {M.}~\bibnamefont
    {Liu}}, \bibinfo {author} {\bibfnamefont {C.-K.}\ \bibnamefont {Duan}},
    \bibinfo {author} {\bibfnamefont {P.~A.}\ \bibnamefont {Tanner}}, \bibinfo
    {author} {\bibfnamefont {C.-G.}\ \bibnamefont {Ma}}, \bibinfo {author}
    {\bibfnamefont {X.}~\bibnamefont {Wei}},\ and\ \bibinfo {author}
    {\bibfnamefont {M.}~\bibnamefont {Yin}},\ }\bibfield  {title} {\bibinfo
    {title} {{Understanding Photoluminescence of
    ${\mathrm{Cs}}_{2}{\mathrm{ZrCl}}_{6}$ Doped with Post-Transition-metal Ions
    Using First-Principles Calculations}},\ }\href
    {https://doi.org/10.1103/PhysRevB.105.195137} {\bibfield  {journal} {\bibinfo
     {journal} {Phys. Rev. B}\ }\textbf {\bibinfo {volume} {105}},\ \bibinfo
    {pages} {195137} (\bibinfo {year} {2022})}\BibitemShut {NoStop}%
  \bibitem [{\citenamefont {Hao}\ and\ \citenamefont {Duan}(2024)}]{2024ic-hrj}%
    \BibitemOpen
    \bibfield  {author} {\bibinfo {author} {\bibfnamefont {R.}~\bibnamefont
    {Hao}}\ and\ \bibinfo {author} {\bibfnamefont {C.-K.}\ \bibnamefont {Duan}},\
    }\bibfield  {title} {\bibinfo {title} {{Unraveling the Photoluminescent
    Properties of Sb-Doped Cd-Based Inorganic Halides: A First-Principles
    Study}},\ }\href {https://doi.org/10.1021/acs.inorgchem.3c04300} {\bibfield
    {journal} {\bibinfo  {journal} {Inorganic Chemistry}\ }\textbf {\bibinfo
    {volume} {63}},\ \bibinfo {pages} {3152} (\bibinfo {year}
    {2024})}\BibitemShut {NoStop}%
  \bibitem [{\citenamefont {Hao}\ \emph {et~al.}(2023)\citenamefont {Hao},
    \citenamefont {Liu}, \citenamefont {Yin},\ and\ \citenamefont
    {Duan}}]{2023jpcc-hrj}%
    \BibitemOpen
    \bibfield  {author} {\bibinfo {author} {\bibfnamefont {R.}~\bibnamefont
    {Hao}}, \bibinfo {author} {\bibfnamefont {M.}~\bibnamefont {Liu}}, \bibinfo
    {author} {\bibfnamefont {M.}~\bibnamefont {Yin}},\ and\ \bibinfo {author}
    {\bibfnamefont {C.-K.}\ \bibnamefont {Duan}},\ }\bibfield  {title} {\bibinfo
    {title} {{Luminescence Mechanism of $ns^{2}$ Ions in Cs$_2$(Sn/Hf)Cl$_6$
    Revealed by First-Principles Calculations}},\ }\href
    {https://doi.org/10.1021/acs.jpcc.2c08071} {\bibfield  {journal} {\bibinfo
    {journal} {Journal of Physical Chemistry C}\ }\textbf {\bibinfo {volume}
    {127}},\ \bibinfo {pages} {3742} (\bibinfo {year} {2023})}\BibitemShut
    {NoStop}%
  \bibitem [{\citenamefont {Oomen}\ \emph {et~al.}(1984)\citenamefont {Oomen},
    \citenamefont {Smit},\ and\ \citenamefont {Blasse}}]{1984cpl-YPO4}%
    \BibitemOpen
    \bibfield  {author} {\bibinfo {author} {\bibfnamefont {E.}~\bibnamefont
    {Oomen}}, \bibinfo {author} {\bibfnamefont {W.}~\bibnamefont {Smit}},\ and\
    \bibinfo {author} {\bibfnamefont {G.}~\bibnamefont {Blasse}},\ }\bibfield
    {title} {\bibinfo {title} {{Jahn-Teller Effect in the Sb$^{3+}$ Emission in
    Zircon-Structured Phosphates}},\ }\href
    {https://doi.org/https://doi.org/10.1016/0009-2614(84)85775-9} {\bibfield
    {journal} {\bibinfo  {journal} {Chemical Physics Letters}\ }\textbf {\bibinfo
    {volume} {112}},\ \bibinfo {pages} {547} (\bibinfo {year}
    {1984})}\BibitemShut {NoStop}%
  \bibitem [{\citenamefont {Oomen}\ \emph {et~al.}(1988)\citenamefont {Oomen},
    \citenamefont {Smit},\ and\ \citenamefont {Blasse}}]{YPO4}%
    \BibitemOpen
    \bibfield  {author} {\bibinfo {author} {\bibfnamefont {E.~W. J.~L.}\
    \bibnamefont {Oomen}}, \bibinfo {author} {\bibfnamefont {W.~M.~A.}\
    \bibnamefont {Smit}},\ and\ \bibinfo {author} {\bibfnamefont
    {G.}~\bibnamefont {Blasse}},\ }\bibfield  {title} {\bibinfo {title}
    {{Jahn-Teller Effect in the Emission and Excitation Spectra of the
    ${\mathrm{Sb}}^{3+}$ Ion in L${\mathrm{PO}}_{4}$ (L=Sc, Lu, Y)}},\ }\href
    {https://doi.org/10.1103/PhysRevB.37.18} {\bibfield  {journal} {\bibinfo
    {journal} {Phys. Rev. B}\ }\textbf {\bibinfo {volume} {37}},\ \bibinfo
    {pages} {18} (\bibinfo {year} {1988})}\BibitemShut {NoStop}%
  \bibitem [{\citenamefont {Bersuker}(2013)}]{2013CR-PJTE}%
    \BibitemOpen
    \bibfield  {author} {\bibinfo {author} {\bibfnamefont {I.~B.}\ \bibnamefont
    {Bersuker}},\ }\bibfield  {title} {\bibinfo {title} {{Pseudo-Jahn-Teller
    Effect-A Two-State Paradigm in Formation, Deformation, and Transformation of
    Molecular Systems and Solids}},\ }\href {https://doi.org/10.1021/cr300279n}
    {\bibfield  {journal} {\bibinfo  {journal} {Chemical Reviews}\ }\textbf
    {\bibinfo {volume} {113}},\ \bibinfo {pages} {1351} (\bibinfo {year}
    {2013})}\BibitemShut {NoStop}%
  \bibitem [{\citenamefont {Bersuker}(2021)}]{2021CR-Bersuker}%
    \BibitemOpen
    \bibfield  {author} {\bibinfo {author} {\bibfnamefont {I.~B.}\ \bibnamefont
    {Bersuker}},\ }\bibfield  {title} {\bibinfo {title} {{Jahn-Teller and
    Pseudo-Jahn-Teller Effects: From Particular Features to General Tools in
    Exploring Molecular and Solid State Properties}},\ }\href
    {https://doi.org/10.1021/acs.chemrev.0c00718} {\bibfield  {journal} {\bibinfo
     {journal} {Chemical Reviews}\ }\textbf {\bibinfo {volume} {121}},\ \bibinfo
    {pages} {1463} (\bibinfo {year} {2021})}\BibitemShut {NoStop}%
  \bibitem [{\citenamefont {Bl\"ochl}(1994)}]{PAW}%
    \BibitemOpen
    \bibfield  {author} {\bibinfo {author} {\bibfnamefont {P.~E.}\ \bibnamefont
    {Bl\"ochl}},\ }\bibfield  {title} {\bibinfo {title} {{Projector
    Augmented-Wave Method}},\ }\href {https://doi.org/10.1103/physrevb.50.17953}
    {\bibfield  {journal} {\bibinfo  {journal} {Physical Review B}\ }\textbf
    {\bibinfo {volume} {50}},\ \bibinfo {pages} {17953} (\bibinfo {year}
    {1994})}\BibitemShut {NoStop}%
  \bibitem [{\citenamefont {Kresse}\ and\ \citenamefont {Hafner}(1993)}]{vasp1}%
    \BibitemOpen
    \bibfield  {author} {\bibinfo {author} {\bibfnamefont {G.}~\bibnamefont
    {Kresse}}\ and\ \bibinfo {author} {\bibfnamefont {J.}~\bibnamefont
    {Hafner}},\ }\bibfield  {title} {\bibinfo {title} {{$Ab$ $initio$ Molecular
    Dynamics for Open-Shell Transition Metals}},\ }\href
    {https://doi.org/10.1103/physrevb.48.13115} {\bibfield  {journal} {\bibinfo
    {journal} {Physical Review B}\ }\textbf {\bibinfo {volume} {48}},\ \bibinfo
    {pages} {13115} (\bibinfo {year} {1993})}\BibitemShut {NoStop}%
  \bibitem [{\citenamefont {Kresse}\ and\ \citenamefont {Hafner}(1994)}]{vasp2}%
    \BibitemOpen
    \bibfield  {author} {\bibinfo {author} {\bibfnamefont {G.}~\bibnamefont
    {Kresse}}\ and\ \bibinfo {author} {\bibfnamefont {J.}~\bibnamefont
    {Hafner}},\ }\bibfield  {title} {\bibinfo {title} {{$Ab$ $initio$
    Molecular-Dynamics Simulation of the Liquid-Metal-Amorphous-Semiconductor
    Transition in Germanium}},\ }\href
    {https://doi.org/10.1103/physrevb.49.14251} {\bibfield  {journal} {\bibinfo
    {journal} {Physical Review B}\ }\textbf {\bibinfo {volume} {49}},\ \bibinfo
    {pages} {14251} (\bibinfo {year} {1994})}\BibitemShut {NoStop}%
  \bibitem [{\citenamefont {Perdew}\ \emph {et~al.}(2008)\citenamefont {Perdew},
    \citenamefont {Ruzsinszky}, \citenamefont {Csonka}, \citenamefont {Vydrov},
    \citenamefont {Scuseria}, \citenamefont {Constantin}, \citenamefont {Zhou},\
    and\ \citenamefont {Burke}}]{PBEsol}%
    \BibitemOpen
    \bibfield  {author} {\bibinfo {author} {\bibfnamefont {J.~P.}\ \bibnamefont
    {Perdew}}, \bibinfo {author} {\bibfnamefont {A.}~\bibnamefont {Ruzsinszky}},
    \bibinfo {author} {\bibfnamefont {G.~I.}\ \bibnamefont {Csonka}}, \bibinfo
    {author} {\bibfnamefont {O.~A.}\ \bibnamefont {Vydrov}}, \bibinfo {author}
    {\bibfnamefont {G.~E.}\ \bibnamefont {Scuseria}}, \bibinfo {author}
    {\bibfnamefont {L.~A.}\ \bibnamefont {Constantin}}, \bibinfo {author}
    {\bibfnamefont {X.}~\bibnamefont {Zhou}},\ and\ \bibinfo {author}
    {\bibfnamefont {K.}~\bibnamefont {Burke}},\ }\bibfield  {title} {\bibinfo
    {title} {{Restoring the Density-Gradient Expansion for Exchange in Solids and
    Surfaces}},\ }\href {https://doi.org/10.1103/PhysRevLett.100.136406}
    {\bibfield  {journal} {\bibinfo  {journal} {Phys. Rev. Lett.}\ }\textbf
    {\bibinfo {volume} {100}},\ \bibinfo {pages} {136406} (\bibinfo {year}
    {2008})}\BibitemShut {NoStop}%
  \bibitem [{\citenamefont {Wang}\ \emph {et~al.}(2021)\citenamefont {Wang},
    \citenamefont {Xu}, \citenamefont {Liu}, \citenamefont {Tang},\ and\
    \citenamefont {Geng}}]{vaspkit}%
    \BibitemOpen
    \bibfield  {author} {\bibinfo {author} {\bibfnamefont {V.}~\bibnamefont
    {Wang}}, \bibinfo {author} {\bibfnamefont {N.}~\bibnamefont {Xu}}, \bibinfo
    {author} {\bibfnamefont {J.-C.}\ \bibnamefont {Liu}}, \bibinfo {author}
    {\bibfnamefont {G.}~\bibnamefont {Tang}},\ and\ \bibinfo {author}
    {\bibfnamefont {W.-T.}\ \bibnamefont {Geng}},\ }\bibfield  {title} {\bibinfo
    {title} {Vaspkit: A user-friendly interface facilitating high-throughput
    computing and analysis using vasp code},\ }\href
    {https://doi.org/10.1016/j.cpc.2021.108033} {\bibfield  {journal} {\bibinfo
    {journal} {Computer Physics Communications}\ }\textbf {\bibinfo {volume}
    {267}},\ \bibinfo {pages} {108033} (\bibinfo {year} {2021})}\BibitemShut
    {NoStop}%
  \bibitem [{\citenamefont {Feng}\ \emph {et~al.}(2021)\citenamefont {Feng},
    \citenamefont {Lou}, \citenamefont {Yin}, \citenamefont {Yeung},
    \citenamefont {Sun},\ and\ \citenamefont {Duan}}]{2021ic-fzy}%
    \BibitemOpen
    \bibfield  {author} {\bibinfo {author} {\bibfnamefont {Z.}~\bibnamefont
    {Feng}}, \bibinfo {author} {\bibfnamefont {B.}~\bibnamefont {Lou}}, \bibinfo
    {author} {\bibfnamefont {M.}~\bibnamefont {Yin}}, \bibinfo {author}
    {\bibfnamefont {Y.-y.}\ \bibnamefont {Yeung}}, \bibinfo {author}
    {\bibfnamefont {H.-T.}\ \bibnamefont {Sun}},\ and\ \bibinfo {author}
    {\bibfnamefont {C.-K.}\ \bibnamefont {Duan}},\ }\bibfield  {title} {\bibinfo
    {title} {{First-Principles Study of Bi$^{3+}$-Related Luminescence and
    Electron and Hole Traps in (Y/Lu/La)PO$_{4}$}},\ }\href
    {https://doi.org/10.1021/acs.inorgchem.0c03217} {\bibfield  {journal}
    {\bibinfo  {journal} {Inorganic Chemistry}\ }\textbf {\bibinfo {volume}
    {60}},\ \bibinfo {pages} {4434} (\bibinfo {year} {2021})}\BibitemShut
    {NoStop}%
  \bibitem [{\citenamefont {Milligan}\ \emph {et~al.}(1982)\citenamefont
    {Milligan}, \citenamefont {Mullica}, \citenamefont {Beall},\ and\
    \citenamefont {Boatner}}]{LPO4}%
    \BibitemOpen
    \bibfield  {author} {\bibinfo {author} {\bibfnamefont {W.}~\bibnamefont
    {Milligan}}, \bibinfo {author} {\bibfnamefont {D.}~\bibnamefont {Mullica}},
    \bibinfo {author} {\bibfnamefont {G.}~\bibnamefont {Beall}},\ and\ \bibinfo
    {author} {\bibfnamefont {L.}~\bibnamefont {Boatner}},\ }\bibfield  {title}
    {\bibinfo {title} {{Structural Investigations of YPO$_{4}$, ScPO$_{4}$, and
    LuPO$_{4}$}},\ }\href
    {https://doi.org/https://doi.org/10.1016/S0020-1693(00)91148-4} {\bibfield
    {journal} {\bibinfo  {journal} {Inorganica Chimica Acta}\ }\textbf {\bibinfo
    {volume} {60}},\ \bibinfo {pages} {39} (\bibinfo {year} {1982})}\BibitemShut
    {NoStop}%
  \bibitem [{\citenamefont {Hinuma}\ \emph {et~al.}(2017)\citenamefont {Hinuma},
    \citenamefont {Pizzi}, \citenamefont {Kumagai}, \citenamefont {Oba},\ and\
    \citenamefont {Tanaka}}]{seekpath}%
    \BibitemOpen
    \bibfield  {author} {\bibinfo {author} {\bibfnamefont {Y.}~\bibnamefont
    {Hinuma}}, \bibinfo {author} {\bibfnamefont {G.}~\bibnamefont {Pizzi}},
    \bibinfo {author} {\bibfnamefont {Y.}~\bibnamefont {Kumagai}}, \bibinfo
    {author} {\bibfnamefont {F.}~\bibnamefont {Oba}},\ and\ \bibinfo {author}
    {\bibfnamefont {I.}~\bibnamefont {Tanaka}},\ }\bibfield  {title} {\bibinfo
    {title} {Band structure diagram paths based on crystallography},\ }\href
    {https://doi.org/https://doi.org/10.1016/j.commatsci.2016.10.015} {\bibfield
    {journal} {\bibinfo  {journal} {Computational Materials Science}\ }\textbf
    {\bibinfo {volume} {128}},\ \bibinfo {pages} {140} (\bibinfo {year}
    {2017})}\BibitemShut {NoStop}%
  \bibitem [{\citenamefont {Hedin}(1965)}]{1965GW-Hedin}%
    \BibitemOpen
    \bibfield  {author} {\bibinfo {author} {\bibfnamefont {L.}~\bibnamefont
    {Hedin}},\ }\bibfield  {title} {\bibinfo {title} {{New Method for Calculating
    the One-Particle Green's Function with Application to the Electron-Gas
    Problem}},\ }\href {https://doi.org/10.1103/PhysRev.139.A796} {\bibfield
    {journal} {\bibinfo  {journal} {Phys. Rev.}\ }\textbf {\bibinfo {volume}
    {139}},\ \bibinfo {pages} {A796} (\bibinfo {year} {1965})}\BibitemShut
    {NoStop}%
  \bibitem [{\citenamefont {Hybertsen}\ and\ \citenamefont
    {Louie}(1986)}]{1986GW0-Hybertsen}%
    \BibitemOpen
    \bibfield  {author} {\bibinfo {author} {\bibfnamefont {M.~S.}\ \bibnamefont
    {Hybertsen}}\ and\ \bibinfo {author} {\bibfnamefont {S.~G.}\ \bibnamefont
    {Louie}},\ }\bibfield  {title} {\bibinfo {title} {{Electron Correlation in
    Semiconductors and Insulators: Band Gaps and Quasiparticle Energies}},\
    }\href {https://doi.org/10.1103/PhysRevB.34.5390} {\bibfield  {journal}
    {\bibinfo  {journal} {Phys. Rev. B}\ }\textbf {\bibinfo {volume} {34}},\
    \bibinfo {pages} {5390} (\bibinfo {year} {1986})}\BibitemShut {NoStop}%
  \bibitem [{sup()}]{supplemental_material}%
    \BibitemOpen
    \href@noop {} {\bibinfo  {journal} {{See Supplemental Material at [URL will
    be inserted by publisher] for detailed information on GW$_0$ band structures,
    demonstration of the absence of finite-size effects, VRBE, bond length of
    [SbO$_{8}$] and barriers without including SOC, which contains
    Refs.~\onlinecite{ScPO4-1997,2009PRB,gap-YPO4,1986GW0-Hybertsen,1984cpl-YPO4,YPO4,lmz-2021jpcc}}}\
    }\BibitemShut {NoStop}%
  \bibitem [{\citenamefont {Henkelman}\ \emph {et~al.}(2000)\citenamefont
    {Henkelman}, \citenamefont {Uberuaga},\ and\ \citenamefont
    {Jónsson}}]{2000JCP-Henkelman1}%
    \BibitemOpen
  \bibfield  {journal} {  }\bibfield  {author} {\bibinfo {author} {\bibfnamefont
    {G.}~\bibnamefont {Henkelman}}, \bibinfo {author} {\bibfnamefont {B.~P.}\
    \bibnamefont {Uberuaga}},\ and\ \bibinfo {author} {\bibfnamefont
    {H.}~\bibnamefont {Jónsson}},\ }\bibfield  {title} {\bibinfo {title} {{A
    Climbing Image Nudged Elastic Band Method for Finding Saddle Points and
    Minimum Energy Paths}},\ }\href {https://doi.org/10.1063/1.1329672}
    {\bibfield  {journal} {\bibinfo  {journal} {The Journal of Chemical Physics}\
    }\textbf {\bibinfo {volume} {113}},\ \bibinfo {pages} {9901–9904} (\bibinfo
    {year} {2000})}\BibitemShut {NoStop}%
  \bibitem [{\citenamefont {Henkelman}\ and\ \citenamefont
    {Jónsson}(2000)}]{2000JCP-Henkelman2}%
    \BibitemOpen
    \bibfield  {author} {\bibinfo {author} {\bibfnamefont {G.}~\bibnamefont
    {Henkelman}}\ and\ \bibinfo {author} {\bibfnamefont {H.}~\bibnamefont
    {Jónsson}},\ }\bibfield  {title} {\bibinfo {title} {{Improved Tangent
    Estimate in the Nudged Elastic Band Method for Finding Minimum Energy Paths
    and Saddle Points}},\ }\href {https://doi.org/10.1063/1.1323224} {\bibfield
    {journal} {\bibinfo  {journal} {The Journal of Chemical Physics}\ }\textbf
    {\bibinfo {volume} {113}},\ \bibinfo {pages} {9978} (\bibinfo {year}
    {2000})}\BibitemShut {NoStop}%
  \bibitem [{\citenamefont {Freysoldt}\ \emph {et~al.}(2014)\citenamefont
    {Freysoldt}, \citenamefont {Grabowski}, \citenamefont {Hickel}, \citenamefont
    {Neugebauer}, \citenamefont {Kresse}, \citenamefont {Janotti},\ and\
    \citenamefont {Van~de Walle}}]{2014rmp-cc}%
    \BibitemOpen
    \bibfield  {author} {\bibinfo {author} {\bibfnamefont {C.}~\bibnamefont
    {Freysoldt}}, \bibinfo {author} {\bibfnamefont {B.}~\bibnamefont
    {Grabowski}}, \bibinfo {author} {\bibfnamefont {T.}~\bibnamefont {Hickel}},
    \bibinfo {author} {\bibfnamefont {J.}~\bibnamefont {Neugebauer}}, \bibinfo
    {author} {\bibfnamefont {G.}~\bibnamefont {Kresse}}, \bibinfo {author}
    {\bibfnamefont {A.}~\bibnamefont {Janotti}},\ and\ \bibinfo {author}
    {\bibfnamefont {C.~G.}\ \bibnamefont {Van~de Walle}},\ }\bibfield  {title}
    {\bibinfo {title} {{First-Principles Calculations for Point Defects in
    Solids}},\ }\href {https://doi.org/10.1103/revmodphys.86.253} {\bibfield
    {journal} {\bibinfo  {journal} {Reviews of Modern Physics}\ }\textbf
    {\bibinfo {volume} {86}},\ \bibinfo {pages} {253} (\bibinfo {year}
    {2014})}\BibitemShut {NoStop}%
  \bibitem [{\citenamefont {Freysoldt}\ \emph {et~al.}(2009)\citenamefont
    {Freysoldt}, \citenamefont {Neugebauer},\ and\ \citenamefont {Van~de
    Walle}}]{2009FNV}%
    \BibitemOpen
    \bibfield  {author} {\bibinfo {author} {\bibfnamefont {C.}~\bibnamefont
    {Freysoldt}}, \bibinfo {author} {\bibfnamefont {J.}~\bibnamefont
    {Neugebauer}},\ and\ \bibinfo {author} {\bibfnamefont {C.~G.}\ \bibnamefont
    {Van~de Walle}},\ }\bibfield  {title} {\bibinfo {title} {{Fully Ab Initio
    Finite-Size Corrections for Charged-Defect Supercell Calculations}},\ }\href
    {https://doi.org/10.1103/PhysRevLett.102.016402} {\bibfield  {journal}
    {\bibinfo  {journal} {Phys. Rev. Lett.}\ }\textbf {\bibinfo {volume} {102}},\
    \bibinfo {pages} {016402} (\bibinfo {year} {2009})}\BibitemShut {NoStop}%
  \bibitem [{\citenamefont {Kumagai}\ and\ \citenamefont {Oba}(2014)}]{2014eFNV}%
    \BibitemOpen
    \bibfield  {author} {\bibinfo {author} {\bibfnamefont {Y.}~\bibnamefont
    {Kumagai}}\ and\ \bibinfo {author} {\bibfnamefont {F.}~\bibnamefont {Oba}},\
    }\bibfield  {title} {\bibinfo {title} {{Electrostatics-Based Finite-Size
    Corrections for First-Principles Point Defect Calculations}},\ }\href
    {https://doi.org/10.1103/PhysRevB.89.195205} {\bibfield  {journal} {\bibinfo
    {journal} {Phys. Rev. B}\ }\textbf {\bibinfo {volume} {89}},\ \bibinfo
    {pages} {195205} (\bibinfo {year} {2014})}\BibitemShut {NoStop}%
  \bibitem [{\citenamefont {Kumagai}\ \emph {et~al.}(2021)\citenamefont
    {Kumagai}, \citenamefont {Tsunoda}, \citenamefont {Takahashi},\ and\
    \citenamefont {Oba}}]{pydefect}%
    \BibitemOpen
    \bibfield  {author} {\bibinfo {author} {\bibfnamefont {Y.}~\bibnamefont
    {Kumagai}}, \bibinfo {author} {\bibfnamefont {N.}~\bibnamefont {Tsunoda}},
    \bibinfo {author} {\bibfnamefont {A.}~\bibnamefont {Takahashi}},\ and\
    \bibinfo {author} {\bibfnamefont {F.}~\bibnamefont {Oba}},\ }\bibfield
    {title} {\bibinfo {title} {{Insights into Oxygen Vacancies from
    High-Throughput First-Principles Calculations}},\ }\href
    {https://doi.org/10.1103/PhysRevMaterials.5.123803} {\bibfield  {journal}
    {\bibinfo  {journal} {Physical Review Materials}\ }\textbf {\bibinfo {volume}
    {5}},\ \bibinfo {pages} {123803} (\bibinfo {year} {2021})}\BibitemShut
    {NoStop}%
  \bibitem [{\citenamefont {Alkauskas}\ \emph {et~al.}(2014)\citenamefont
    {Alkauskas}, \citenamefont {Yan},\ and\ \citenamefont {Van~de
    Walle}}]{2014prb-cc}%
    \BibitemOpen
    \bibfield  {author} {\bibinfo {author} {\bibfnamefont {A.}~\bibnamefont
    {Alkauskas}}, \bibinfo {author} {\bibfnamefont {Q.}~\bibnamefont {Yan}},\
    and\ \bibinfo {author} {\bibfnamefont {C.~G.}\ \bibnamefont {Van~de Walle}},\
    }\bibfield  {title} {\bibinfo {title} {{First-Principles Theory of
    Nonradiative Carrier Capture via Multiphonon Emission}},\ }\href
    {https://doi.org/10.1103/PhysRevB.90.075202} {\bibfield  {journal} {\bibinfo
    {journal} {Phys. Rev. B}\ }\textbf {\bibinfo {volume} {90}},\ \bibinfo
    {pages} {075202} (\bibinfo {year} {2014})}\BibitemShut {NoStop}%
  \bibitem [{\citenamefont {Jones}\ and\ \citenamefont {Gunnarsson}(1989)}]{SCF}%
    \BibitemOpen
    \bibfield  {author} {\bibinfo {author} {\bibfnamefont {R.~O.}\ \bibnamefont
    {Jones}}\ and\ \bibinfo {author} {\bibfnamefont {O.}~\bibnamefont
    {Gunnarsson}},\ }\bibfield  {title} {\bibinfo {title} {{The Density
    Functional Formalism, Its Applications and Prospects}},\ }\href
    {https://doi.org/10.1103/RevModPhys.61.689} {\bibfield  {journal} {\bibinfo
    {journal} {Rev. Mod. Phys.}\ }\textbf {\bibinfo {volume} {61}},\ \bibinfo
    {pages} {689} (\bibinfo {year} {1989})}\BibitemShut {NoStop}%
  \bibitem [{\citenamefont {Bersuker}(2006)}]{2006Bersuker}%
    \BibitemOpen
    \bibfield  {author} {\bibinfo {author} {\bibfnamefont {I.~B.}\ \bibnamefont
    {Bersuker}},\ }\href {https://api.semanticscholar.org/CorpusID:94776995}
    {\emph {\bibinfo {title} {{The Jahn-Teller Effect}}}}\ (\bibinfo  {publisher}
    {Cambridge University Press},\ \bibinfo {address} {Cambridge},\ \bibinfo
    {year} {2006})\BibitemShut {NoStop}%
  \bibitem [{\citenamefont {Sio}\ and\ \citenamefont
    {Giustino}(2023)}]{2022Natphy}%
    \BibitemOpen
    \bibfield  {author} {\bibinfo {author} {\bibfnamefont {W.~H.}\ \bibnamefont
    {Sio}}\ and\ \bibinfo {author} {\bibfnamefont {F.}~\bibnamefont {Giustino}},\
    }\bibfield  {title} {\bibinfo {title} {{Polarons in Two-Dimensional Atomic
    Crystals}},\ }\href {https://api.semanticscholar.org/CorpusID:256854096}
    {\bibfield  {journal} {\bibinfo  {journal} {Nature Physics}\ }\textbf
    {\bibinfo {volume} {19}},\ \bibinfo {pages} {629} (\bibinfo {year}
    {2023})}\BibitemShut {NoStop}%
  \bibitem [{\citenamefont {Dai}\ \emph {et~al.}(2024{\natexlab{a}})\citenamefont
    {Dai}, \citenamefont {Lian}, \citenamefont {Lafuente-Bartolome},\ and\
    \citenamefont {Giustino}}]{2024PRL-exciton}%
    \BibitemOpen
    \bibfield  {author} {\bibinfo {author} {\bibfnamefont {Z.}~\bibnamefont
    {Dai}}, \bibinfo {author} {\bibfnamefont {C.}~\bibnamefont {Lian}}, \bibinfo
    {author} {\bibfnamefont {J.}~\bibnamefont {Lafuente-Bartolome}},\ and\
    \bibinfo {author} {\bibfnamefont {F.}~\bibnamefont {Giustino}},\ }\bibfield
    {title} {\bibinfo {title} {{Excitonic Polarons and Self-Trapped Excitons from
    First-Principles Exciton-Phonon Couplings}},\ }\href
    {https://doi.org/10.1103/PhysRevLett.132.036902} {\bibfield  {journal}
    {\bibinfo  {journal} {Phys. Rev. Lett.}\ }\textbf {\bibinfo {volume} {132}},\
    \bibinfo {pages} {036902} (\bibinfo {year} {2024}{\natexlab{a}})}\BibitemShut
    {NoStop}%
  \bibitem [{\citenamefont {Dai}\ \emph {et~al.}(2024{\natexlab{b}})\citenamefont
    {Dai}, \citenamefont {Lian}, \citenamefont {Lafuente-Bartolome},\ and\
    \citenamefont {Giustino}}]{2024PRB-exciton}%
    \BibitemOpen
    \bibfield  {author} {\bibinfo {author} {\bibfnamefont {Z.}~\bibnamefont
    {Dai}}, \bibinfo {author} {\bibfnamefont {C.}~\bibnamefont {Lian}}, \bibinfo
    {author} {\bibfnamefont {J.}~\bibnamefont {Lafuente-Bartolome}},\ and\
    \bibinfo {author} {\bibfnamefont {F.}~\bibnamefont {Giustino}},\ }\bibfield
    {title} {\bibinfo {title} {{Theory of Excitonic Polarons: From Models to
    First-Principles Calculations}},\ }\href
    {https://doi.org/10.1103/PhysRevB.109.045202} {\bibfield  {journal} {\bibinfo
     {journal} {Phys. Rev. B}\ }\textbf {\bibinfo {volume} {109}},\ \bibinfo
    {pages} {045202} (\bibinfo {year} {2024}{\natexlab{b}})}\BibitemShut
    {NoStop}%
  \bibitem [{\citenamefont {Ham}(1965)}]{1965PRB-Ham}%
    \BibitemOpen
    \bibfield  {author} {\bibinfo {author} {\bibfnamefont {F.~S.}\ \bibnamefont
    {Ham}},\ }\bibfield  {title} {\bibinfo {title} {{Dynamical Jahn-Teller Effect
    in Paramagnetic Resonance Spectra: Orbital Reduction Factors and Partial
    Quenching of Spin-Orbit Interaction}},\ }\href
    {https://doi.org/10.1103/PhysRev.138.A1727} {\bibfield  {journal} {\bibinfo
    {journal} {Phys. Rev.}\ }\textbf {\bibinfo {volume} {138}},\ \bibinfo {pages}
    {A1727} (\bibinfo {year} {1965})}\BibitemShut {NoStop}%
  \bibitem [{\citenamefont {Streltsov}\ and\ \citenamefont
    {Khomskii}(2020)}]{2020PRX-JTE-SOC}%
    \BibitemOpen
    \bibfield  {author} {\bibinfo {author} {\bibfnamefont {S.~V.}\ \bibnamefont
    {Streltsov}}\ and\ \bibinfo {author} {\bibfnamefont {D.~I.}\ \bibnamefont
    {Khomskii}},\ }\bibfield  {title} {\bibinfo {title} {{Jahn-Teller Effect and
    Spin-Orbit Coupling: Friends or Foes?}},\ }\href
    {https://doi.org/10.1103/PhysRevX.10.031043} {\bibfield  {journal} {\bibinfo
    {journal} {Phys. Rev. X}\ }\textbf {\bibinfo {volume} {10}},\ \bibinfo
    {pages} {031043} (\bibinfo {year} {2020})}\BibitemShut {NoStop}%
  \bibitem [{\citenamefont {Lyu}\ and\ \citenamefont
    {Dorenbos}(2018)}]{2018jmcc-Bi}%
    \BibitemOpen
    \bibfield  {author} {\bibinfo {author} {\bibfnamefont {T.}~\bibnamefont
    {Lyu}}\ and\ \bibinfo {author} {\bibfnamefont {P.}~\bibnamefont {Dorenbos}},\
    }\bibfield  {title} {\bibinfo {title} {{Bi$^{3+}$ Acting Both as An Electron
    and as A Hole Trap in La-, Y-, and LuPO$_{4}$}},\ }\href
    {https://api.semanticscholar.org/CorpusID:139206948} {\bibfield  {journal}
    {\bibinfo  {journal} {Journal of Materials Chemistry C}\ }\textbf {\bibinfo
    {volume} {6}},\ \bibinfo {pages} {6240} (\bibinfo {year} {2018})}\BibitemShut
    {NoStop}%
  \bibitem [{\citenamefont {Awater}\ \emph {et~al.}(2017)\citenamefont {Awater},
    \citenamefont {Niemeijer-Berghuijs},\ and\ \citenamefont
    {Dorenbos}}]{2017OM-Bi}%
    \BibitemOpen
    \bibfield  {author} {\bibinfo {author} {\bibfnamefont {R.~H.}\ \bibnamefont
    {Awater}}, \bibinfo {author} {\bibfnamefont {L.~C.}\ \bibnamefont
    {Niemeijer-Berghuijs}},\ and\ \bibinfo {author} {\bibfnamefont
    {P.}~\bibnamefont {Dorenbos}},\ }\bibfield  {title} {\bibinfo {title}
    {{Luminescence and Charge Carrier Trapping in YPO$_{4}$:Bi}},\ }\href
    {https://api.semanticscholar.org/CorpusID:100133599} {\bibfield  {journal}
    {\bibinfo  {journal} {Optical Materials}\ }\textbf {\bibinfo {volume} {66}},\
    \bibinfo {pages} {351} (\bibinfo {year} {2017})}\BibitemShut {NoStop}%
  \bibitem [{\citenamefont {Srivastava}\ and\ \citenamefont
    {Camardello}(2015)}]{2015OM-Bi}%
    \BibitemOpen
    \bibfield  {author} {\bibinfo {author} {\bibfnamefont {A.~M.}\ \bibnamefont
    {Srivastava}}\ and\ \bibinfo {author} {\bibfnamefont {S.~J.}\ \bibnamefont
    {Camardello}},\ }\bibfield  {title} {\bibinfo {title} {{Concentration
    Dependence of the Bi$^{3+}$ Luminescence in LnPO$_{4}$ (Ln = Y$^{3+}$,
    Lu$^{3+}$)}},\ }\href {https://api.semanticscholar.org/CorpusID:96812729}
    {\bibfield  {journal} {\bibinfo  {journal} {Optical Materials}\ }\textbf
    {\bibinfo {volume} {39}},\ \bibinfo {pages} {130} (\bibinfo {year}
    {2015})}\BibitemShut {NoStop}%
  \bibitem [{\citenamefont {Trukhin}\ and\ \citenamefont
    {Boatner}(1997)}]{ScPO4-1997}%
    \BibitemOpen
    \bibfield  {author} {\bibinfo {author} {\bibfnamefont {A.~N.}\ \bibnamefont
    {Trukhin}}\ and\ \bibinfo {author} {\bibfnamefont {L.~A.}\ \bibnamefont
    {Boatner}},\ }\bibfield  {title} {\bibinfo {title} {{Electronic Structure of
    ScPO$_4$ Single Crystals: Optical and Photoelectric Properties}},\ }\href
    {https://api.semanticscholar.org/CorpusID:138666798} {\bibfield  {journal}
    {\bibinfo  {journal} {Materials Science Forum}\ }\textbf {\bibinfo {volume}
    {239-241}},\ \bibinfo {pages} {573 } (\bibinfo {year} {1997})}\BibitemShut
    {NoStop}%
  \bibitem [{\citenamefont {Krumpel}\ \emph {et~al.}(2009)\citenamefont
    {Krumpel}, \citenamefont {Bos}, \citenamefont {Bessi\`ere}, \citenamefont
    {van~der Kolk},\ and\ \citenamefont {Dorenbos}}]{2009PRB}%
    \BibitemOpen
    \bibfield  {author} {\bibinfo {author} {\bibfnamefont {A.~H.}\ \bibnamefont
    {Krumpel}}, \bibinfo {author} {\bibfnamefont {A.~J.~J.}\ \bibnamefont {Bos}},
    \bibinfo {author} {\bibfnamefont {A.}~\bibnamefont {Bessi\`ere}}, \bibinfo
    {author} {\bibfnamefont {E.}~\bibnamefont {van~der Kolk}},\ and\ \bibinfo
    {author} {\bibfnamefont {P.}~\bibnamefont {Dorenbos}},\ }\bibfield  {title}
    {\bibinfo {title} {{Controlled Electron and Hole Trapping in
    ${\text{YPO}}_{4}:{\text{Ce}}^{3+},{\text{Ln}}^{3+}$ and
    ${\text{LuPO}}_{4}:{\text{Ce}}^{3+},{\text{Ln}}^{3+}$ ($\text{Ln}=\text{Sm}$,
    Dy, Ho, Er, Tm)}},\ }\href {https://doi.org/10.1103/PhysRevB.80.085103}
    {\bibfield  {journal} {\bibinfo  {journal} {Phys. Rev. B}\ }\textbf {\bibinfo
    {volume} {80}},\ \bibinfo {pages} {085103} (\bibinfo {year}
    {2009})}\BibitemShut {NoStop}%
  \bibitem [{\citenamefont {Bos}\ \emph {et~al.}(2008)\citenamefont {Bos},
    \citenamefont {Dorenbos}, \citenamefont {Bessi{\`e}re},\ and\ \citenamefont
    {Viana}}]{gap-YPO4}%
    \BibitemOpen
    \bibfield  {author} {\bibinfo {author} {\bibfnamefont {A.~J.~J.}\
    \bibnamefont {Bos}}, \bibinfo {author} {\bibfnamefont {P.}~\bibnamefont
    {Dorenbos}}, \bibinfo {author} {\bibfnamefont {A.}~\bibnamefont
    {Bessi{\`e}re}},\ and\ \bibinfo {author} {\bibfnamefont {B.}~\bibnamefont
    {Viana}},\ }\bibfield  {title} {\bibinfo {title} {{Lanthanide Energy Levels
    in YPO$_{4}$}},\ }\href {https://api.semanticscholar.org/CorpusID:97900964}
    {\bibfield  {journal} {\bibinfo  {journal} {Radiation Measurements}\ }\textbf
    {\bibinfo {volume} {43}},\ \bibinfo {pages} {222} (\bibinfo {year}
    {2008})}\BibitemShut {NoStop}%
  \end{thebibliography}


\end{document}